\documentclass[10pt]{article}
\usepackage{amsmath,amssymb,amsthm,amsopn,mathrsfs,mathdots,mathtools}
\usepackage{graphicx,verbatim,array,multicol,courier,dsfont,soul}
\usepackage[pdftex,bookmarks,colorlinks]{hyperref}
\usepackage{fullpage,anysize,relsize,color}
\usepackage[margin=1in]{geometry}
\usepackage[comma]{natbib}
\usepackage[section]{placeins}
\usepackage[labelformat=simple]{subfig}
\usepackage[applemac]{inputenc}
\usepackage{booktabs}
\usepackage[makeroom]{cancel}
\usepackage{enumerate}
\usepackage{enumitem,float}

\newlist{inparaenum}{enumerate}{2}
\setlist[inparaenum,1]{label=(\roman*)}
\setlist[inparaenum,2]{label=(\roman{inparaenumi}\emph{\alph*})}

\newcommand{\der}{{\mathrm d}}

\def\cD{\mathcal{D}}
\def\natural{{\mathbb N}}

\def\der{\text{d}}

\def\real{{\mathbb R}}
\def\nat{{\mathbb N}}

\newtheorem{theo}{Theorem}[section]

\newtheorem{prop}{Proposition}[section]

\newtheorem{lem}{Lemma}

\hypersetup{colorlinks,%
citecolor=blue,%
filecolor=green,%
linkcolor=red,%
urlcolor=magenta,%
}

\begin{document}

\title{Extremal properties of the univariate extended skew-normal distribution}

\author{B. Beranger\footnote{School of Mathematics and Statistics, University of New South Wales, Sydney, Australia.}\;\,\footnote{Communicating Author: {\tt B.Beranger@unsw.edu.au}},\; S. A. Padoan\footnote{Department of Decision Sciences, Bocconi University of Milan, Italy.
},\; Y. Xu$^*$\; and\; S. A. Sisson$^*$
}

\date{}

\maketitle
\begin{abstract}
We consider the extremal properties of the highly flexible univariate extended skew-normal distribution.
We derive the well-known Mills' inequalities and Mills' ratio for the extended skew-normal distribution and establish the asymptotic 
extreme-value distribution for the maximum of samples drawn from this distribution.
\\

\noindent Keywords: Domain of attraction; Generalized extreme-value distribution; Gumbel distribution; Normalising constants; Rates of convergence; Von Mises function.
\end{abstract}
%

%%%%%%%%%%%%%%%%%%%%%%%%%%%%%%%%%%%%%%%%%%%%%
%
% SECTION 1
%
%%%%%%%%%%%%%%%%%%%%%%%%%%%%%%%%%%%%%%%%%%%%%
%
\section{Introduction and background}\label{sec:intro}

The skew-normal and related families are classes of asymmetric probability distributions that include the normal distribution as a 
special case \citep{azzalini1985, azzalini2014}.
In recent years they have received increasing interest from the scientific community because in many applications data is frequently incompatible with symmetric distributions, such as the normal or elliptical distributions.

In this work we focus on the univariate extended skew-normal distribution 
\citep[Ch 5.]{arellano2010, azzalini2014}. Precisely, 
a random variable $X$ follows an extended skew-normal distribution \citep{arellano2010}, denoted 
as $X\sim ESN(\mu, \omega, \alpha, \tau)$, if its probability density function (pdf) is given by
\begin{equation}
\label{eq:pdfESN}
\phi(x; \mu, \omega, \alpha, \tau) = 
\frac{\phi(x;\mu,\omega)}{\Phi\left(\tau/\sqrt{1+ \alpha^2} \right)} 
\Phi(\alpha z + \tau), 
\quad x\in \real,
\end{equation}
where $\phi(x;\mu,\omega)$ is a univariate normal pdf with 
mean $\mu \in \real$ and standard deviation $\omega>0$, $z = \omega^{-1}(x-\mu)$, 
and $\Phi(\cdot)$ is the standard univariate normal cumulative distribution function (cdf).
The parameters $\alpha \in \real$ and $\tau \in \real$ are known as  the slant and  extension parameters, respectively, and they control the nature of density deviations away from normality. When $\tau = 0$ the extended skew-normal distribution reduces to the skew-normal $SN(\mu,\omega,\alpha)$, and when both $\tau=0$ and  $\alpha= 0$ the normal $N(\mu,\omega)$ distribution is obtained. 
Without loss of generality, we work with location and scale standardised distributions throughout, i.e.~$\mu=0$ and $\omega=1$
which we compactly denote as $ESN(\alpha,\tau)$. For further visual and presentational clarity  we write  distributional parameters in the subscript of the pdf and cdf so that e.g. $\phi(x; \alpha,\tau)=\phi_{\alpha,\tau}(x)$ and $\Phi(x;\alpha,\tau)=\Phi_{\alpha,\tau}(x)$.

Our contribution concerns the derivation of the extremal properties of the univariate extended skew-normal distribution.
Specifically, we obtain the well-known Mills' inequalities and ratio \citep{mills1926},
and as a result, derive the asymptotic extreme-value distribution for the maximum of an extended skew-normal random sample, for large 
sample sizes. The speed at which the sample distribution converges to its limiting case is also determined.

We briefly recall the cornerstone result of univariate extreme-value theory. For each $n\in\natural$, let $X_1,\ldots,X_n$ be a series of independent and identically distributed ({\em iid}) univariate random variables
with a continuous distribution function $F$ defined on $\real$.
Define, the ($n$-partial) sample maximum by  $M_n=\max(X_1,\ldots,X_n)$.
If there is a sequence of normalising constants $a_n>0$ and $b_n\in\real$ such that 
\begin{equation}
\label{eq:limDist}
\lim_{n\to\infty}
\Pr \left( \frac{M_n -b_n}{a_n} \leq x \right) 
= \lim_{n\to\infty} F^n(a_n x + b_n)=G(x),
\end{equation}
for all continuity points $x\in\real$ of $G$, then this limiting distribution must be a member of the
Generalized Extreme-Value (GEV) family of distributions, denoted by $G_\gamma$ where
$\gamma\in\real$ is the tail-index parameter \cite[Section~2.1]{beirlant2004}. 
Specifically, members of the GEV class of distributions are: the standard Gumbel $G_0(x)$, Fr\'{e}chet $G_\alpha(x)=G_{1/\gamma}((x-1)/\gamma)$ for $\gamma>0$ and negative Weibull $G_\beta(x)=G_{-1/\gamma}(-(x+1)/\gamma)$ 
for $\gamma<0$. When the limit \eqref{eq:limDist} holds we say that
$F$ is in the maximum-domain of attraction of $G_\gamma$, in symbols $F \in \cD(G_\gamma)$.
One well known result is that a necessary and sufficient condition for $F \in \cD(G_\gamma)$ is that 
$F$ is a Von Mises function, see~\citet[Ch. 1]{resnick1987} and ~\citet[Proposition 1.4]{resnick1987} for the special case when $F \in \cD(G_0)$.

The remainder of this paper is organised as follows.
In Section~\ref{sec:esn_family} we firstly derive  Mills' inequalities and ratio, and then the extreme-value 
distribution for the sample maximum and the convergence rate of its sample distribution. 
Throughout, all proofs are provided in the Appendix.

%%%%%%%%%%%%%%%%%%%%%%%%%%%%%%%%%%%%%%%%%%%%
%
% SECTION 3: EXTREMAL PROPERTIES OF THE ESN DISTRIBUTION
%
%%%%%%%%%%%%%%%%%%%%%%%%%%%%%%%%%%%%%%%%%%%%
%
\section{Extremes of extended skew-normal random samples}\label{sec:esn_family}
%

%%%%%%%%%%%%%%%%%%%%%%%%%%%%%%%%%%%%%%%%%%%%
%
% SECTION 3.1: UNIVARIATE CASE
%
%%%%%%%%%%%%%%%%%%%%%%%%%%%%%%%%%%%%%%%%%%%%
%

\citet{mills1926} established the following results (Mill's inequalities and Mills' ratio) for the standard normal distribution:
\begin{eqnarray}
 x^{-1} (1+x^{-2})^{-1} &<& \frac{1-\Phi(x)}{\phi(x)} < x^{-1},\quad x>0\label{eqn:millsInequality},\\
\frac{1-\Phi(x)}{\phi(x)} &\approx& x^{-1},\quad x\to\infty\label{eqn:millsRatio},
\end{eqnarray}
where (\ref{eqn:millsRatio}) is obtained from (\ref{eqn:millsInequality}) for large $x$. Mills' ratio can be used to establish the normalising constants $a_n$ and $b_n$ in (\ref{eq:limDist}) following Proposition 1.1 in \citet{resnick1987}.
\citet{liao2014} derived Mills' inequalities and ratio for the 
skew-normal distribution, from which  (\ref{eqn:millsInequality}) and (\ref{eqn:millsRatio})
may be recovered by setting $\alpha=0$.
Here we require more general results. The following two propositions derive  Mills' inequalities (Proposition \ref{prop:MillsIneq}) and ratio (Proposition \ref{prop:MillsRatio}) for the
extended skew-normal distribution. It follows that the results in
\citet{liao2014} can be obtained from these by setting $\tau=0$.
\begin{prop}[Mills' Inequality]
\label{prop:MillsIneq}
Let $X\sim ESN (\alpha,\tau)$ where $\alpha,\tau\in\real$. 
For each $x\in\real$ and $\alpha,\tau\in\real$, define $x_{\alpha,\tau}:=\alpha x +\tau$
and $\bar{\alpha}=(1+\alpha^2)^{1/2}$.
For any $x>0$ we have
$$
L_{\alpha,\tau}(x) < \frac{1-\Phi_{\alpha,\tau}(x)}{\phi_{\alpha,\tau}(x)} < U_{\alpha,\tau}(x),
$$
where the upper and lower bounds are given as follows:
\begin{inparaenum}
\item \label{en:first_cond_palpha} when $\alpha\geq 0$
\begin{inparaenum}
\item \label{en:first_cond_palpha_parg} and when $x_{\alpha,\tau}>0$, then 
$$
\begin{array}{ccc}
L_{\alpha,\tau}(x)=x^{-1} \left(1+x^{-2}\right)^{-1} &
\textrm{and} &
U_{\alpha,\tau}(x)=x^{-1} \left(1-\frac{\phi(x_{\alpha,\tau})}{x_{\alpha,\tau}}\right)^{-1},
\end{array}
$$
\item \label{en:first_cond_palpha_narg} and when $ x_{\alpha,\tau}<0$, then 
$$
\begin{array}{ccc}
L_{\alpha,\tau}(x)=x^{-1}\left(1+x^{-2}\right)^{-1} &
\textrm{and} &
U_{\alpha,\tau}(x)=x^{-1}\left(-\frac{x_{\alpha,\tau}^2+1}{x_{\alpha,\tau}\phi(x_{\alpha,\tau})}\right),
\end{array}
$$
\end{inparaenum}
\item \label{en:second_cond_nalpha} when $\alpha<0$
\begin{inparaenum}
\item \label{en:second_cond_nalpha_parg} and when $x_{\alpha,\tau}>0$ and $x+\alpha x_{\alpha,\tau}>0$, then
\begin{align*}
L_{\alpha,\tau}(x)& = x^{-1} \left(1+x^{-2}\right)^{-1} \left(1+ \frac{\alpha}{\bar{\alpha}^2x+\alpha\tau}
\frac{\phi(x_{\alpha,\tau}) x_{\alpha,\tau}}{x_{\alpha,\tau}-\phi(x_{\alpha,\tau})}\right),\\
U_{\alpha,\tau}(x)&=x^{-1} \left(1+\frac{\alpha(\bar{\alpha}^2x+\alpha\tau)}
{(\bar{\alpha}^2x+\alpha\tau)^2+\bar{\alpha}^2}
\frac{\phi(x_{\alpha,\tau})(\bar{\alpha}^2x+\alpha\tau)}
{x_{\alpha,\tau}^2+1-x_{\alpha,\tau}\phi(x_{\alpha,\tau})}
\right),
\end{align*}
\item \label{en:second_cond_nalpha_mixed} and when $x_{\alpha,\tau}>0$ and $x+\alpha x_{\alpha,\tau}<0$, then
\begin{align*}
L_{\alpha,\tau}(x)& = x^{-1} \left(1+x^{-2}\right)^{-1}
\left(
1+\frac{ 
\frac{\alpha}{\bar{\alpha}} 
\phi\left(\bar{\alpha}x+\frac{\alpha\tau}{\bar{\alpha}}\right)^{-1}
+\frac{\alpha\{\bar{\alpha}^2x+\alpha\tau\}}
{\{\bar{\alpha}^2x+\alpha\tau\}^2+\bar{\alpha}^2}
}
{ \phi(x_{\alpha,\tau})^{-1} - x_{\alpha,\tau}^{-1}} 
\right),\\
U_{\alpha,\tau}(x)& = x^{-1} \left(
1+\frac{
\frac{\alpha}{\bar{\alpha}}
\phi\left(\bar{\alpha}x+\frac{\alpha\tau}{\bar{\alpha}}\right)^{-1}
+\frac{\alpha}{\bar{\alpha}^2x+\alpha\tau}
}
{
\phi(x_{\alpha,\tau})^{-1}
-\frac{x_{\alpha,\tau}}{x_{\alpha,\tau}^2+1}
}
\right),
\end{align*}
\item \label{en:second_cond_nalpha_mixed} and when $x_{\alpha,\tau}<0$, then
\begin{align*}
L_{\alpha,\tau}(x) &= x^{-1} \left(1+x^{-2}\right)^{-1} 
\left\{
1-\frac{\alpha x_{\alpha,\tau}}
{\bar{\alpha}^2x+\alpha\tau}
\left(1+\frac{1}{x_{\alpha,\tau}^2}\right)
\right\},\\
U_{\alpha,\tau}(x) &= x^{-1} \left(
1-\frac{\alpha x_{\alpha,\tau}}
{\bar{\alpha}^2x+\alpha\tau}
\left(1+\frac{\bar{\alpha}^2}{(\bar{\alpha}^2x+\alpha\tau)^2}\right)^{-1}
\right).
\end{align*}
\end{inparaenum}
\end{inparaenum}
\end{prop}

\begin{prop}[Mills' Ratio]
\label{prop:MillsRatio}  
Let $X\sim ESN (\alpha,\tau)$ with $\alpha,\tau\in\real$.
Then, from Proposition \ref{prop:MillsIneq}, as $x \rightarrow \infty$ we have
$$
\frac{1-\Phi_{\alpha,\tau}(x)}{\phi_{\alpha,\tau}(x)} \approx
\left\{
\begin{array}{cc}
x^{-1} & \alpha \geq 0 \\
\left\{ (\alpha^2 + 1)x + \alpha \tau \right\}^{-1} & \alpha < 0.
\end{array}
\right.
$$
\end{prop}

Given Mills' ratio for the extended skew-normal derived in Proposition \ref{prop:MillsRatio}, Proposition \ref{prop:DomAtt} demonstrates sufficient conditions on the survival function $1-\Phi_{\alpha,\tau}(x)$ to conclude that the extended skew-normal distribution $\Phi_{\alpha,\tau}$ is both a Von Mises function, and is in the maximum domain of attraction of the Gumbel distribution  $\Phi_{\alpha,\tau}\in\cD(G_0)$, regardless of whether $\alpha\geq 0$ or $\alpha<0$.

\begin{prop}[Gumbel Domain of Attraction]
\label{prop:DomAtt}
Let $X\sim ESN (\alpha,\tau)$ with $\alpha,\tau\in\real$. For $x\to\infty$, the survival function $1-\Phi_{\alpha,\tau}(x)$ can be written as
$$
1-\Phi_{\alpha,\tau}(x)=c(x) \exp \left( - \int_{1}^x \frac{g(v)}{f(v)} \der v\right).
$$
In particular, when $\alpha \geq 0$, then as $x\to\infty$
\begin{align*}
c(x)&\to
\frac{1}{\Phi(\tau / \bar{\alpha})\sqrt{2\pi e}} >0,\\
g(x) &= 1 + x^{-2}\to 1,\\
f(x) &= x^{-1}>0, \quad f'(x) = - x^{-2} \to 0,
\end{align*}
where $\bar{\alpha}=(1+\alpha^2)^{1/2}$, whereas when $\alpha < 0$, 
without loss of generality, assume that $\alpha+\tau<0$ and $\bar{\alpha}^2+\alpha\tau>0$, then as $x\to\infty$
\begin{align*}
c(x) &\to\frac{-\exp\left( - \frac{1 + (\alpha + \tau)^2}{2} \right)}
{2 \pi \Phi\left(\tau / \bar{\alpha}\right) 
(\alpha (\alpha + \tau) + 1)(\alpha + \tau)} >0,\\
g(x) &= 1 + \frac{\bar{\alpha}^2}{\left( \bar{\alpha}^2x + \alpha\tau \right)^2} 
+ \frac{\alpha}{(\alpha x + \tau)\left( \bar{\alpha}^2x + \alpha\tau \right)}
\to 1,\\
f(x) &= \frac{1}{ \bar{\alpha}^2 x +\alpha\tau} >0, 
\quad f'(x) = \frac{-\bar{\alpha}^2}{\left(\bar{\alpha}^2x +\alpha\tau\right)^2}\to 0.
\end{align*}
As a consequence, $\Phi_{\alpha,\tau}$ is both a Von  Mises function and $\Phi_{\alpha,\tau}\in \cD(G_0)$.
\end{prop}
From Proposition~\ref{prop:DomAtt} and in combination with Proposition 1.1 in \citet{resnick1987}, it follows that 
the normalising constants $a_n>0$ and $b_n\in\real$ in \eqref{eq:limDist} can then be identified
through the standard identities
\begin{equation}
\label{eq:AnBn}
\begin{array}{ccc}
1-\Phi_{\alpha,\tau}(b_n) = n^{-1}\quad\mbox{ and }\quad
& & a_n = f(b_n),
\end{array}
\end{equation}
where $f$ is given in Proposition~\ref{prop:DomAtt}.
For practical purposes, it is usually more convenient to identify alternative normalising constants with a closed-form expression.
In general terms, it is well-known that if there are normalising constants $\alpha_n>0$ and
$\beta_n\in\real$, different from $a_n$ and $b_n$, such that $F^n(\alpha_nx + \beta_n)\to \tilde{G}(x)$ converges to a non-degenerate limit $\tilde{G}(x)$ as $n\to\infty$, 
then $\tilde{G}$ is equal to $G$ as given in \eqref{eq:limDist} apart from some modification of the scale and location parameters
\citep[e.g,][Proposition 0.2]{resnick1987}, which does not qualitatively change the tail behavour.
In  Proposition \ref{prop:AlphaBeta} below, we provide some alternative normalising constants $\alpha_n,\beta_n$ with a closed-form expression,
that satisfy the conditions $\alpha_n/ a_n \rightarrow 1$ and $(\beta_n - b_n)a_n \rightarrow 0$,  as $n \rightarrow \infty$ \citep[see][Theorem~1.2.3]{leadbetter1983}.
This therefore implies that the limiting distribution for the normalised sample maximum 
is still a standard Gumbel distribution.
%

%
% PROPOSITION ALPHA_n & BETA_n
%
\begin{prop}[Alternative Normalising Constants]
\label{prop:AlphaBeta}
Let $X_1,\ldots,X_n$ be a series of iid random variables with $X_i\sim ESN (\alpha,\tau)$ for $i=1,\ldots,n$ with $\alpha,\tau\in\real$.
Define $M_n = \max(X_1,\ldots,X_n)$ and define the  normalising constants
\begin{align*}
\alpha_n &= \ell_{n,0}^{-1}\quad \mbox{and}\quad \beta_n = \ell_{n,0}-\frac{\ln(2\sqrt{\pi})+(1/2)\ln\ln n  +\ln\Phi\left(\tau/\bar{\alpha}\right)}{\ell_{n,0}} \qquad\qquad\qquad\;\; \mbox{if }\alpha\geq 0,\\
\alpha_n &= \ell_{n,\alpha}^{-1}\quad \mbox{and}\quad \beta_n = \ell_{n,\alpha}
-\frac{2\ln(2\sqrt{\pi}|\alpha|)+\ln\ln n +\ln\Phi\left(\tau/\bar{\alpha}\right)-\tau^2/2}
{2\ell_{n,\alpha}}-\frac{\alpha\tau}{\bar{\alpha}^2} \quad \mbox{if }\alpha < 0,
\end{align*}
where $\ell_{n,\alpha}:=\sqrt{2\ln n (1+\alpha^2)}$ and $\bar{\alpha}=(1 + \alpha^2)^{1/2}$. Then
\begin{equation*}
\label{eq:limDistAlphaBeta}
\lim_{n\to\infty}\Pr \left( \frac{M_n - \beta_n}{\alpha_n}  \leq x \right)=G_0(x),\quad x\in\real.
\end{equation*}
\end{prop}

In the presence of competing normalising constants, a natural question to ask is
whether the rate of convergence of $\Phi^n_{\alpha,\tau}$ to
$G_0$ as $n\rightarrow\infty$, differs substantially when
the normalising constants $(\alpha_n,\beta_n)$ in Proposition \ref{prop:AlphaBeta} are considered in the place of 
$(a_n,b_n)$ defined by \eqref{eq:AnBn}. Theorem \ref{theo:RateAlphaBeta} establishes the rate of convergence for each sequence of 
normalising constants. 
%
% THEOREM RATE ALPHA_N BETA_N
%
\begin{theo}[Convergence Rate to Gumbel Limit]
\label{theo:RateAlphaBeta}
Let $X\sim ESN (\alpha,\tau)$ with $\alpha,\tau\in\real$.
For the normalising constants $\alpha_n$, $\beta_n$ defined in 
Proposition~\ref{prop:AlphaBeta} we have
$$
\{\Phi_{\alpha,\tau}^n (\alpha_n x + \beta_n) - G_0(x)\}
\approx \frac{G_0(x)e^{-x}}{c} \frac{(\ln\ln n)^2}{\ln n}
\quad \mbox{as }n \rightarrow \infty,
$$
where $c=16$ when $\alpha\geq0$ and $c=4$ when $\alpha<0$.
For the normalising constants $a_n$, $b_n$ defined in \eqref{eq:AnBn} we have
$$
\lim_{n\to\infty}b_n^2 \left[ 
b_n^2 \left\{ \Phi_{\alpha,\tau}^n(a_n x + b_n) - G_0(x) \right\}
- \kappa(x) G_0(x)
\right]
= 
\left(
\omega(x) + \frac{\kappa^2(x)}{2}
\right) G_0(x),
$$
where
$$
\kappa(x) = \frac{x^2 + 2x}{2} e^{-x}\quad\mbox{and}\quad \omega(x) = -\frac{1}{8} \left( x^4 + 4x^3 + 8x^2 + 16x \right) e^{-x}
$$
when $\alpha\geq0$, while
$$
\kappa(x) = \frac{x^2 + 4x}{2(1+\alpha^2)} e^{-x}\quad\mbox{and}\quad \omega(x) =
- \frac{\alpha^2 (1 + \alpha^2)^2}{8} 
\left\{ (1 + 3 \alpha^2) 16 x + \alpha^2 (x^4 + 8x^3 + 24 x^2) \right\} e^{-x}
$$
when $\alpha < 0$.
\end{theo}

From Theorem~\ref{theo:RateAlphaBeta} it follows that when the sample maximum is normalised by $(\alpha_n, \beta_n)$
and $(a_n, b_n)$, then the rates of convergence to the standard the Gumbel distribution are
of order $(\ln\ln n)^2/\ln n$ and $1/\ln n$, respectively. That is, the rate of convergence is slower for $(\alpha_n,\beta_n)$, balancing the advantage of the closed-form expression.

\section*{Acknowledgments}

We thank two anonymous referees for having carefully read the first version of this manuscript and for their helpful comments that have contributed to improving the presentation of this article. 
BB and SAS are supported by the Australian Centre of Excellence in Mathematical and Statistical Frontiers (ACEMS; CE140100049) and by the Australian Research Council Discovery Projects Scheme (FT170100079). SAP is supported by the Bocconi Institute for Data Science and Analytics (BIDSA).

%%%%%%%%%%%%%%%%%%%%%%%%%%%%%%%%%%%%%%%%%%%%%%%%%%
%
% APPENDIX SECTION
%
%
%%%%%%%%%%%%%%%%%%%%%%%%%%%%%%%%%%%%%%%%%%%%%%%%%%
\appendix
\section{Proofs}\label{sec:appendix}
%

%%%%%%%%%%%% PROOF PROPOSITION 1 %%%%%%%%%%%%%%%%%%
%
%
%
\subsection{Proof of Proposition~\ref{prop:MillsIneq}}
Define $x_{\alpha,\tau}:=\alpha x+\tau$ for every $x,\alpha,\tau\in\real$.
From \eqref{eq:pdfESN} we can write 
$$
\frac{1-\Phi_{\alpha,\tau}(x)}{\phi_{\alpha,\tau}(x)}
= \frac{\int_{x}^{+\infty} \phi(t) \Phi(t_{\alpha,\tau}) \der t}{ \phi(x) \Phi(x_{\alpha,\tau})}
= \frac{\int_{x}^{+\infty} \Phi (t_{\alpha,\tau})e^{-t^2/2} \der t}
{\Phi(x_{\alpha,\tau})e^{-x^2/2}}.
$$
For $x>0$, using integration by parts gives
\begin{align}
\label{eq:Ineq1Prop1}
\frac{1}{x^2} \int_{x}^{+\infty} \Phi(t_{\alpha,\tau}) e^{-t^2/2} \der t 
&> \int_{x}^{+\infty} \frac{\Phi(t_{\alpha,\tau})}{t^2}  e^{-t^2/2} \der t
\nonumber\\
&=\frac{\Phi(x_{\alpha,\tau})}{x}  e^{-x^2/2} -\int_{x}^{+\infty} \Phi(t_{\alpha,\tau}) e^{-t^2/2} \der t 
+\alpha \int_{x}^{+\infty} \frac{e^{-t^2/2}}{t}
\phi(t_{\alpha,\tau}) \der t,
\end{align}
from which it follows that
\begin{equation}\label{eq:Ineq2Prop1}
\left(1+\frac{1}{x^2}\right) \int_{x}^{+\infty} \Phi(t_{\alpha,\tau}) 
e^{-t^2/2} \der t 
> \frac{\Phi(x_{\alpha,\tau})}{x} e^{-x^2/2}
+\alpha \int_{x}^{+\infty} \frac{e^{-t^2/2}}{t}
\phi(t_{\alpha,\tau}) \der t.
\end{equation}
We study the behaviour of the inequality \eqref{eq:Ineq2Prop1} conditional on the sign of $\alpha$.
	
\begin{inparaenum}
\item \label{en:first_cond_palpha} When $\alpha>0$, by \eqref{eq:Ineq2Prop1} we obtain
\begin{align*}
\left(1+\frac{1}{x^2}\right) \int_{x}^{+\infty} \Phi(t_{\alpha,\tau}) 
e^{-t^2/2} \der t 
&> x^{-1}\Phi(x_{\alpha,\tau}) e^{-x^2/2},
\end{align*}
and therefore we derive the lower bound
\begin{align*}
L_{\alpha,\tau}(x)= x^{-1} \left(1+\frac{1}{x^2}\right)^{-1} < 
 \frac{ \int_{x}^{+\infty} \Phi(t_{\alpha,\tau}) e^{-t^2/2} \der t}
{\Phi(x_{\alpha,\tau}) e^{-x^2/2}}.
\end{align*}
Since $\int_{x}^{+\infty}	t^{-2} \Phi(t_{\alpha,\tau}) 
e^{-t^2/2} \der t >0$, then from \eqref{eq:Ineq1Prop1} we obtain
\begin{align}
\int_{x}^{+\infty} \Phi(t_{\alpha,\tau}) e^{-t^2/2} \der t 
&< \frac{\Phi(x_{\alpha,\tau})}{x} e^{-\frac{x^2}{2}} \der t 
+ \alpha \int_{x}^{+\infty} t^{-1} e^{-\frac{t^2}{2}} \phi(t_{\alpha,\tau}) \der t 
\nonumber\\
&< \frac{\Phi(x_{\alpha,\tau})}{x} e^{-\frac{x^2}{2}}
\left(1+
\frac{\alpha\int_{x}^{+\infty}  e^{-\frac{t^2}{2}} \phi(t_{\alpha,\tau}) \der t}
{\Phi(x_{\alpha,\tau}) e^{-x^2/2}}
\right)
< \frac{\Phi(x_{\alpha,\tau})}{x} e^{-\frac{x^2}{2}} 
\left(1+\frac{\alpha\int_{x}^{+\infty} \phi(t_{\alpha,\tau}) \der t}
{\Phi(x_{\alpha,\tau})} \right) \nonumber\\
&= \frac{\Phi(x_{\alpha,\tau})}{x} e^{-\frac{x^2}{2}} 
\left(1+\frac{1-\Phi(x_{\alpha,\tau})}
{\Phi(x_{\alpha,\tau})} \right)
= x^{-1}e^{-\frac{x^2}{2}}.\nonumber
\end{align}
Then, we have
\begin{equation}\label{eq:Ineq3Prop1}
\frac{1-\Phi_{\alpha,\tau}(x)}{\phi_{\alpha,\tau}(x)}\leq x^{-1}\Phi^{-1}(x_{\alpha,\tau}).
\end{equation}
We now also need to consider the sign of $x_{\alpha,\tau}$. 
\begin{inparaenum}
\item \label{en:first_cond_palpha_parg} When $x_{\alpha,\tau} > 0$, by 
\eqref{eqn:millsInequality}  
we have
$
1-\Phi(x_{\alpha,\tau}) < \phi(x_{\alpha,\tau})/ x_{\alpha,\tau},
$
which implies
$$
\Phi(x_{\alpha,\tau})^{-1} < \left(1-\frac{\phi(x_{\alpha,\tau})}{x_{\alpha,\tau}}\right)^{-1}.
$$
Then, substituting the above inequality into \eqref{eq:Ineq3Prop1} we obtain the upper bound
$$
\frac{\int_{x}^{+\infty}\Phi(t_{\alpha,\tau}) e^{-t^2/2} \der t}
{\Phi(x_{\alpha,\tau}) e^{-x^2/2}}
< x^{-1} \left(1-\frac{\phi(x_{\alpha,\tau})}{x_{\alpha,\tau}}\right)^{-1} =U_{\alpha,\tau}(x);
$$

\item \label{en:first_cond_palpha_narg} When $x_{\alpha,\tau}<0$, by inequality \eqref{eqn:millsInequality} we have that
$
\Phi(x_{\alpha,\tau})^{-1} < - \left(x_{\alpha,\tau}^2+1 \right)/\{ x_{\alpha,\tau}\phi(x_{\alpha,\tau})\},
$
and substituting this into \eqref{eq:Ineq3Prop1} gives the upper bound 
$$
\frac{ \int_{x}^{+\infty} \Phi(t_{\alpha,\tau}) e^{-t^2/2} \der t}
{\Phi(x_{\alpha,\tau}) e^{-x^2/2}}
< x^{-1} \left(-\frac{(x_{\alpha,\tau})^2+1}{x_{\alpha,\tau}\phi(x_{\alpha,\tau})}\right) =U_{\alpha,\tau}(x).
$$
\end{inparaenum}
\item \label{en:second_cond_nalpha} When $\alpha < 0 $, using the  property
$$
\phi \left(\frac{x-\mu_1}{\sigma_1} \right) 
\phi \left(\frac{x-\mu_2}{\sigma_2} \right)
= 
\phi \left(
\frac{x \sqrt{\sigma_1^2+\sigma_2^2}}{\sigma_1 \sigma_2} 
-\frac{ \mu_1 \sigma_2^2 + \mu_2 \sigma_1^2}
{ \sqrt{\sigma_1^2+\sigma_2^2} \sigma_1 \sigma_2}
\right)
\phi \left( \frac{\mu_1 - \mu_2}{\sqrt{\sigma_1^2+\sigma_2^2}}\right),
$$
we can write
$$
\exp\left(-\frac{(t_{\alpha,\tau})^2+t^2}{2}\right)
= 2\pi \phi \left(\frac{t+\tau \alpha^{-1}}{\alpha^{-1}}\right) \phi(t)
= 2\pi \phi \left(\bar{\alpha} t+\frac{\tau \alpha^{-1}}{\bar{\alpha}}\right) 
\phi\left(\frac{\tau}{\bar{\alpha}} \right),
$$
where $\bar{\alpha}=\sqrt{1+\alpha^2}$, which then gives
\begin{align}
\label{eq:Ineq4Prop1}
\alpha \int_{x}^{+\infty} e^{-t^2/2} \phi(t_{\alpha,\tau}) \der t
&=\alpha \sqrt{2\pi} \int_{x}^{+\infty}
\phi \left(t \bar{\alpha} +\frac{\tau \alpha^{-1}}{\bar{\alpha}}\right) 
\phi\left(\frac{\tau}{\bar{\alpha}} \right) \der t \nonumber\\
&=\frac{\alpha}{\bar{\alpha}}e^{-\tau^2/(2\bar{\alpha}^2)}
\left\{1-\Phi\left(\bar{\alpha}x+\frac{\alpha\tau}{\bar{\alpha}}\right)\right\}.
\end{align}
Hence, starting from \eqref{eq:Ineq2Prop1}, a lower bound is of the form
\begin{align}
\label{eq:Ineq5Prop1}
 \left(1+\frac{1}{x^2}\right) \int_{x}^{+\infty} \Phi(t_{\alpha,\tau}) 
e^{-t^2/2} \der t
&> \frac{\Phi(x_{\alpha,\tau})}{x} e^{-x^2/2}
\left(1+
\frac{\alpha\int_{x}^{+\infty} t^{-1} e^{-t^2/2}
\phi(t_{\alpha,\tau}) \der t}
{x^{-1}\Phi(x_{\alpha,\tau}) e^{-x^2/2}}
\right) \nonumber\\
&> \frac{\Phi(x_{\alpha,\tau})}{x}  e^{-x^2/2}
\left(1+
\frac{\alpha\int_{x}^{+\infty} e^{-t^2/2} \phi(t_{\alpha,\tau}) \der t}
{\Phi(x_{\alpha,\tau}) e^{-x^2/2}}
\right) \nonumber\\
&= \frac{\Phi(x_{\alpha,\tau})}{x} e^{-x^2/2} 
\left(1+
\frac{\frac{\alpha}{\bar{\alpha}}e^{-\frac{\tau^2}{2\bar{\alpha}^2}}
\left\{1-\Phi\left(\bar{\alpha}x+\frac{\alpha\tau}{\bar{\alpha}}\right)\right\}}
{\Phi(x_{\alpha,\tau}) e^{-x^2/2}}\right).
\end{align}

Similarly, an upper bound is given by
\begin{align*}
x \int_{x}^{+\infty} \Phi(t_{\alpha,\tau}) e^{-t^2/2} \der t
< \int_{x}^{+\infty} t\Phi(t_{\alpha,\tau}) e^{-t^2/2} \der t
= \Phi(x_{\alpha,\tau}) e^{-x^2/2}
+ \int_{x}^{+\infty} \alpha\phi(t_{\alpha,\tau})  e^{-t^2/2} \der t,
\end{align*}
and hence, using \eqref{eq:Ineq4Prop1} we obtain
\begin{align}
\label{eq:Ineq6Prop1}
 \int_{x}^{+\infty}\Phi(t_{\alpha,\tau})e^{-t^2/2} \der t
&< x^{-1} \Phi(x_{\alpha,\tau}) e^{-x^2/2}
\left(1+
\frac{\int_{x}^{+\infty}\alpha\phi(t_{\alpha,\tau})  e^{-t^2/2} \der t}
{\Phi(x_{\alpha,\tau}) e^{-x^2/2}}\right) \nonumber\\
&=x^{-1} \Phi(x_{\alpha,\tau}) e^{-x^2/2}
\left(1+
\frac{\frac{\alpha}{\bar{\alpha}}e^{-\tau^2/(2\bar{\alpha}^2)}
\left\{1-\Phi\left(\bar{\alpha}x+\frac{\alpha\tau}{\bar{\alpha}}\right)\right\}}
{\Phi(x_{\alpha,\tau}) e^{-x^2/2}}\right).
\end{align}
As before, we also need to consider the sign of $x_{\alpha,\tau}$.
\begin{inparaenum}
\item \label{en:second_cond_nalpha_parg} When $x_{\alpha,\tau}>0$ and $x+\alpha x_{\alpha,\tau}>0$ then 
we have $x\bar{\alpha}+(\alpha\tau)/\bar{\alpha}>0$. Therefore, by \eqref{eqn:millsInequality} we obtain
$
1-\Phi(x_{\alpha,\tau}) < \phi(x_{\alpha,\tau})/ x_{\alpha,\tau},
$
and using the equality
\begin{equation}\label{eq:Eq1Prop1}
e^{-\tau^2/(2\bar{\alpha}^2)}
\phi \left( \bar{\alpha}x + \frac{\alpha \tau}{\bar{\alpha}} \right)
= e^{-x^2/2}
\phi \left( x_{\alpha,\tau} \right)
\end{equation}
we have that
\begin{align*}
& \frac{ \frac{\alpha}{\bar{\alpha}} e^{-\tau^2/(2\bar{\alpha}^2)}
\left\{1-\Phi\left(\bar{\alpha}x + \frac{\alpha\tau}{\bar{\alpha}}\right)\right\}}
{\Phi(x_{\alpha,\tau}) e^{-x^2/2}} 
 > \frac{ \alpha \left(\bar{\alpha}^2x+\alpha\tau\right)^{-1}}
{1/\phi(x_{\alpha,\tau}) - 1/x_{\alpha,\tau}}.
\end{align*}
Thus, substituting the above equality into \eqref{eq:Ineq5Prop1} leads to the desired 
lower bound $L_{\alpha,\tau}(x)$.
Further, from \eqref{eqn:millsInequality} we also have
$
\Phi(x_{\alpha,\tau})<1-\phi(x_{\alpha,\tau})x_{\alpha,\tau}/(x_{\alpha,\tau}^2+1),
$
which, combined with \eqref{eq:Eq1Prop1}, gives
\begin{align*}
\frac{\frac{\alpha}{\bar{\alpha}}e^{-\tau^2/(2\bar{\alpha}^2)}
\left\{1-\Phi\left(\bar{\alpha}x+\frac{\alpha\tau}{\bar{\alpha}}\right)\right\}}
{\Phi(x_{\alpha,\tau}) e^{-x^2/2}}
< \frac{
\frac{\alpha(\bar{\alpha}^2x+\alpha\tau)}
{(\bar{\alpha}^2x+\alpha\tau)^2+\bar{\alpha}^2}
}
{\phi(x_{\alpha,\tau})^{-1}-\frac{x_{\alpha,\tau}}{x_{\alpha,\tau}^2+1}}.
\end{align*}
Applying the above inequality to \eqref{eq:Ineq6Prop1} leads to the upper bound $U_{\alpha,\tau}(x)$.
\item When $x_{\alpha,\tau}>0$ and $x+\alpha x_{\alpha,\tau}<0$ then we have
$\bar{\alpha}x+(\alpha\tau)/\bar{\alpha}<0$. Therefore, by \eqref{eqn:millsInequality} we obtain
\begin{align*}
1-\Phi\left(\bar{\alpha}x + \alpha\tau/\bar{\alpha}\right) 
< 1+\frac{\bar{\alpha}}{\bar{\alpha}^2x+\alpha\tau}
\frac{\phi\left(\bar{\alpha}x+\alpha\tau/\bar{\alpha}\right)}
{1+ 1/\left(\bar{\alpha}x+\alpha\tau/\bar{\alpha}\right)^{2}},
\end{align*}
which, combined with \eqref{eq:Eq1Prop1}, gives
\begin{align*}
\frac{\frac{\alpha}{\bar{\alpha}}e^{-\tau^2/(2\bar{\alpha})}
\left\{1-\Phi\left(\bar{\alpha}x+\alpha\tau/\bar{\alpha}\right)\right\}}
{\Phi(x_{\alpha,\tau}) e^{-x^2/2}}
> \frac{\frac{\alpha}{\bar{\alpha}} 
\phi\left(\bar{\alpha}x+\frac{\alpha\tau}{\bar{\alpha}}\right)^{-1}+
\frac{\alpha(\bar{\alpha}^2x+\alpha\tau)}
{(\bar{\alpha}^2x+\alpha\tau)^2+\bar{\alpha}^2}}
{1/\phi(x_{\alpha,\tau}) - 1/x_{\alpha,\tau}}.
\end{align*}
Together with \eqref{eq:Ineq5Prop1} the above inequality leads to the desired 
lower bound $L_{\alpha,\tau}(x)$.
By \eqref{eqn:millsInequality} we also have
$$
1-\Phi\left(\bar{\alpha}x+\alpha\tau/\bar{\alpha}\right)
> 1+\frac{\phi\left(\bar{\alpha}x+\alpha\tau/\bar{\alpha}\right)}{\bar{\alpha}x+\alpha\tau/\bar{\alpha}},
$$
and from this it follows that
\begin{align*}
\frac{\frac{\alpha}{\bar{\alpha}}e^{-\tau^2/(2\bar{\alpha}^2)}
\left\{1-\Phi\left(\bar{\alpha}x+\alpha\tau/\bar{\alpha}\right)\right\}}
{\Phi(x_{\alpha,\tau}) e^{-x^2/2}}
< \frac{
\frac{\alpha}{\bar{\alpha}}e^{-\tau^2/(2\bar{\alpha}^2)}
\left(1+
\frac{\phi\left(\bar{\alpha}x+\frac{\alpha\tau}{\bar{\alpha}}\right)}
{\bar{\alpha}x+\frac{\alpha\tau}{\bar{\alpha}}}
\right)}
{e^{-x^2/2}\left\{1- x_{\alpha,\tau}\phi(x_{\alpha,\tau})/(x_{\alpha,\tau}^2+1)\right\}}.
\end{align*}
Together with \eqref{eq:Ineq6Prop1} the above inequality leads to the 
upper bound $U_{\alpha,\tau}(x)$.

\item \label{en:second_cond_nalpha_narg} When $x_{\alpha,\tau}<0 $,
since $\alpha/\bar{\alpha}<0$  and 
$\bar{\alpha}x+(\alpha\tau)\bar{\alpha}>0$, 
then by \eqref{eqn:millsInequality} we have
$$
1-\Phi\left(\bar{\alpha}x+\alpha\tau/\bar{\alpha}\right)
	<\frac{\phi\left(\bar{\alpha}x+\alpha\tau/\bar{\alpha}\right)}{\bar{\alpha}x+\alpha\tau/\bar{\alpha}},
$$
and thus
\begin{align}
\label{eq:Ineq7Prop1}
\frac{\alpha}{\bar{\alpha}} e^{-\tau^2/(2\bar{\alpha}^2)}
\left\{1-\Phi\left(\bar{\alpha}x+\frac{\alpha\tau}{\bar{\alpha}}\right)\right\}
> \frac{\alpha}
{\bar{\alpha}\left(\bar{\alpha}x
+\frac{\alpha\tau}{\bar{\alpha}}\right)}
e^{-\tau^2/(2\bar{\alpha}^2)} 
\phi\left(\bar{\alpha}x+\frac{\alpha\tau}{\bar{\alpha}}\right).
\end{align}
Furthermore, by (\ref{eqn:millsInequality}) we also have
$$
\Phi(x_{\alpha,\tau}) e^{-x^2/2}
>-\frac{x_{\alpha,\tau}}{x_{\alpha,\tau}^2+1}\phi(x_{\alpha,\tau})
e^{-x^2/2},
$$
and combined with \eqref{eq:Eq1Prop1}, \eqref{eq:Ineq7Prop1} then becomes
$$
\frac{
\frac{\alpha}{\bar{\alpha}} e^{-\tau^2/(2\bar{\alpha}^2)}
\left\{1-\Phi\left(\bar{\alpha}x+\alpha\tau/\bar{\alpha}\right)\right\}}
{\Phi(x_{\alpha,\tau}) e^{-x^2/2}}
>-\frac{\alpha(x_{\alpha,\tau}^2+1)}{(\bar{\alpha}^2x+\alpha\tau)x_{\alpha,\tau}}.
$$
Together with \eqref{eq:Ineq5Prop1} the above inequality provides the 
lower bound $L_{\alpha,\tau}(x)$.
Again applying \eqref{eqn:millsInequality} we have
$
\Phi(x_{\alpha,\tau}) < \phi(x_{\alpha,\tau})/x_{\alpha,\tau}
$
and
$$
1-\Phi\left(\bar{\alpha}x+\alpha\tau/\bar{\alpha}\right)
>\frac{\phi\left(\bar{\alpha}x+\alpha\tau/\bar{\alpha}\right)}
{\bar{\alpha}x+\alpha\tau/\bar{\alpha}}
\left(1+\left(\bar{\alpha}x+\alpha\tau/\bar{\alpha}\right)^{-2}\right)^{-1}.
$$
Using $\alpha/\bar{\alpha}<0$ and (\ref{eqn:millsInequality}) gives
$$
 \frac{\frac{\alpha}{\bar{\alpha}} e^{-\tau^2/(2\bar{\alpha}^2)}
 \left\{1-\Phi\left(\bar{\alpha}x+\alpha\tau/\bar{\alpha}\right)\right\}}
 {\Phi(x_{\alpha,\tau}) e^{-x^2/2}}
 <-\frac{\alpha x_{\alpha,\tau}(x+\alpha x_{\alpha,\tau})}
 {(\bar{\alpha}^2x+\alpha\tau)^2+\alpha^2+1},	
 $$
which together with \eqref{eq:Ineq5Prop1} provides the upper bound $U_{\alpha,\tau}(x)$.
\end{inparaenum}
\end{inparaenum}
%

%
%
%
%%%%%%%%%%%% END PROOF PROPOSITION 1 %%%%%%%%%%%%%%%%

%%%%%%%%%%%% START PROOF PROPOSITION 2 %%%%%%%%%%%%%%%
%
%
%
\subsection{Proof of Proposition~\ref{prop:MillsRatio}}

Let $x_{\alpha,\tau}:=\alpha x+\tau$ and $\bar{\alpha}=(1+\alpha^2)^{1/2}$.
First, note that when $\alpha>0$, then $x_{\alpha,\tau}$ becomes positive when
$x \rightarrow \infty$, regardless of the value of $\tau$. From Proposition~\ref{prop:MillsIneq}
and noting that  
$$
\lim_{x\rightarrow\infty} \left( 1 - \frac{\phi(x_{\alpha,\tau})}{x_{\alpha,\tau}}\right)^{-1}=1
$$
we can obtain
$$
L_{\alpha,\tau}(x) =\frac{1}{x + x^{-1}} \approx \frac{1}{x},\quad U_{\alpha,\tau}(x) \approx \frac{1}{x}.
$$
Conversely, when $\alpha < 0 $, then $x_{\alpha,\tau}$ becomes negative when
$x \rightarrow \infty$, regardless of the value of $\tau$. From Proposition~\ref{prop:MillsIneq}
we have that the dominating term for the lower and upper bounds is
$$
x^{-1}\left(1-\frac{\alpha x_{\alpha,\tau} }{\bar{\alpha}^2x+\alpha\tau}\right).
$$
As a consequence,
$$
L_{\alpha,\tau}(x) \approx \frac{1}{\bar{\alpha}^2x+\alpha\tau},\quad U_{\alpha,\tau}(x) \approx \frac{1}{\bar{\alpha}^2x+\alpha\tau}.
$$
%
%
%
%%%%%%%%%%%% END PROOF PROPOSITION 2 %%%%%%%%%%%%%%%%

%%%%%%%%%%%% START PROOF PROPOSITION 3 %%%%%%%%%%%%%%%
%
%
%
%\subsection{Proof of Proposition~3.3}\label{app:DomAtt}
\subsection{Proof of Proposition~\ref{prop:DomAtt}}\label{sec:DomAtt}
Define $x_{\alpha,\tau}:=\alpha x+\tau$ and and $\bar{\alpha}=(1+\alpha^2)^{1/2}$ for every $x,\alpha,\tau\in\real$.
When $\alpha\geq0$, by Proposition~\ref{prop:MillsRatio} as $x\to\infty$ we have
\begin{align*}
1-\Phi_{\alpha, \tau}(x)
\approx \frac{\Phi( x_{\alpha,\tau})}{\Phi(\tau / \bar{\alpha})}\frac{e^{-x^2/2}}{x\sqrt{2\pi}}
&= \frac{\Phi(x_{\alpha,\tau})}{\Phi(\tau / \bar{\alpha})} \frac{1}{\sqrt{2\pi e}}
e^{\left(\frac{1}{2} - \ln x -\frac{x^2}{2}\right)}\\
&= \frac{\Phi( x_{\alpha,\tau})}{\Phi(\tau / \bar{\alpha})} \frac{1}{\sqrt{2\pi e}}
\exp\left(- \int_1^x \frac{\frac{t^2 + 1}{t^2}}{\frac{1}{t}} \der t\right)\\
&=c(x)\exp\left(- \int_1^x \frac{g(t)}{f(t)} \der t\right).
\end{align*}
It follows that
$$
\lim_{x\to\infty}c(x)=\frac{1}{\Phi(\tau / \bar{\alpha})\sqrt{2\pi e}},\quad
\lim_{x\to\infty} g(x)=1,\quad \lim_{x\to\infty} f'(x)=0.
$$
When $\alpha <0$, from Proposition~\ref{prop:MillsRatio}  as $x\to\infty$ we have
\begin{align*}
1- \Phi_{\alpha, \tau}(x)&\approx \frac{-\phi(x) \phi(x_{\alpha,\tau})}
{ \left( \bar{\alpha}^2x + \alpha\tau \right)x_{\alpha,\tau} \Phi(\tau / \bar{\alpha})}\\
&=\frac{e^{- \frac{1 + (\alpha + \tau)^2}{2}}}
{2\pi \Phi(\tau / \bar{\alpha})} \exp\left\{ -\frac{1}{2} (x^2 -1)( \alpha^2 +1) - \alpha \tau (x-1)
- \ln(\alpha(x_{\alpha,\tau})+x) -\ln (-x_{\alpha,\tau})\right\}\\
&= \frac{e^{- \frac{1 + (\alpha + \tau)^2}{2}}}
{2\pi \Phi(\tau / \bar{\alpha})}
\exp \left\{ - \ln(\alpha(\alpha +\tau)+1) -\ln( -(\alpha + \tau) ) -\int_1^x \left(\bar{\alpha}^2 t + \lambda\tau 
+ \frac{ \bar{\alpha}^2 }{\alpha t_{\alpha,\tau}+t} 
+ \frac{\alpha}{t_{\alpha,\tau}}\right) \der t
\right\}\\
&= \frac{-e^{- \frac{1 + (\alpha + \tau)^2}{2}}(\alpha + \tau)^{-1}}
{2\pi \Phi(\tau / \bar{\alpha}) \{\alpha(\alpha +\tau)+1\}}
\exp \left\{ -\int_1^x 
\frac{
1 
+ \frac{\alpha^2 + 1}{\left( \bar{\alpha}^2 t+\alpha\tau\right)^2}
+ \frac{\alpha}{(\alpha t + \tau)( \bar{\alpha}^2 t+\alpha\tau)}
}
{
\frac{1}{ \bar{\alpha}^2 t+\alpha\tau}
}
 \der t
\right\}\\
&=c(x)\exp\left(- \int_1^x \frac{g(t)}{f(t)} \der t\right),
\end{align*}
assuming $\bar{\alpha} + \alpha \tau >0$ and $\alpha + \tau <0$.
It follows that 
$$
\lim_{x\to\infty}c(x)=\frac{-e^{- \frac{1 + (\alpha + \tau)^2}{2}} (\alpha + \tau)^{-1}}
{2\pi \Phi(\tau / \bar{\alpha}) \{\alpha(\alpha +\tau)+1\} }>0,
\quad \lim_{x\to\infty}g(x)=1,\quad \lim_{x\to\infty} f'(x)=0.
$$
Therefore, by Proposition 1.1(a) and Corollary 1.7 in \citet{resnick1987}  we have that
$\Phi_{\alpha,\tau}$ is a Von Mises function and $\Phi_{\alpha,\tau}\in\cD(G_0)$.
%
%
%
%
%%%%%%%%%%%% END PROOF PROPOSITION 3 %%%%%%%%%%%%%%%

%%%%%%%%%%%% START PROOF PROPOSITION 4 %%%%%%%%%%%%%%
%
%
%
\subsection{Proof of Proposition~3.4}
\label{sec:AlphaBeta}

Recall that for brevity we write $\ell_{n,\alpha}=\sqrt{2(1+\alpha^2)\ln n}$ and
$\bar{\alpha}=(1+\alpha^2)^{1/2}$ for
any $n\in\nat$ and $\alpha\in\real$.
By Proposition~\ref{prop:DomAtt} we know that $\Phi_{\alpha,\tau}\in\cD(G_0)$ and therefore by Proposition 1.1 in \citet{resnick1987} we have that the normalising constants $a_n>0$ and $b_n\in\real$
can be obtained by solving the equation in \eqref{eq:AnBn}.
Here we derive some approximations for
$a_n$ and $b_n$. We distinguish two cases. First, we consider $\alpha\geq0$. 
By Proposition~\ref{prop:DomAtt} the second equation of \eqref{eq:AnBn} gives $a_n=1/b_n$ 
and by Proposition~\ref{prop:MillsRatio} the left hand-side term of the first equation can be approximated as $1-\Phi_{\alpha,\tau}(b_n)\approx\phi_{\alpha,\tau}(b_n)/ b_n$
as $n\to\infty$, and so through tail equivalence we can focus on the equation
$
n\phi_{\alpha,\tau}(b_n)= b_n.
$
Taking the logarithm on both sides we
obtain
\begin{equation}\label{eq:proxy_eq}
\ln n -\ln b_n -\frac{1}{2} \ln 2\pi - \frac{b_n^2}{2} + \ln \Phi(\alpha b_n + \tau)
- \ln \Phi(\tau / \bar{\alpha}) = 0.
\end{equation}
Dividing~\eqref{eq:proxy_eq} by $b_n^2$ gives $b_n=\ell_{n,0}+o(1)$ and
\begin{equation}
\label{eq:logUn}
\log b_n = \frac{1}{2} (\ln 2 + \ln\ln n) + o(1).
\end{equation}
We set $\alpha_n=1/\ell_{n,0}$. 
Using the fact that $\Phi(\alpha b_n + \tau) \rightarrow 1$ as $n\rightarrow\infty$
and plugging \eqref{eq:logUn} in \eqref{eq:proxy_eq} we obtain
\begin{align*}
b_n =\ell_{n,0}  
- \frac{1/2\ln\ln n + \ln(2\sqrt{\pi}) +\ln 2\Phi(\tau / \bar{\alpha}) }
{\ell_{n,0}} 
+ o\left( 1/\ell_{n,0}\right)
= \beta_n + o(\alpha_n).
\end{align*}
In the second case we assume $\alpha<0$.
By Proposition~\ref{prop:MillsRatio} the left hand-side term of the first equation can be approximated by $\phi_{\alpha,\tau}(b_n)/ \{(1+\alpha^2)b_n+\alpha\tau\}$
as $n\to\infty$, and so through tail equivalence we can focus on the equation
$
n\phi_{\alpha,\tau}(b_n)= (1+\alpha^2)b_n+\alpha\tau.
$
By taking the logarithm on both sides 
and noting that from Proposition~\ref{prop:MillsIneq}
$$
\ln \Phi(\alpha b_n +\tau)\approx -\frac{1}{2}\ln 2\pi-\frac{(\alpha b_n+\tau)^2}{2}-\ln (-(\alpha b_n+\tau)),\quad n\to\infty,
$$
then  we obtain
\begin{equation}\label{eq:proxy_eq_2}
\ln n -\ln \{(1+\alpha^2)b_n+\alpha\tau\} - \ln 2\pi \alpha\tau b_n-\frac{\tau^2}{2}- \frac{b_n^2\bar{\alpha}^2}{2} + \ln \Phi\{-(\alpha b_n + \tau)\}
- \ln \Phi(\tau / \bar{\alpha}) = 0.
\end{equation}
Dividing~\eqref{eq:proxy_eq_2} by $b_n^2$ then gives $b_n=\ell_{n,\alpha}+o(1)$.
We set $\alpha_n=1/\ell_{n,\alpha}$.
Plugging $b_n$ in \eqref{eq:proxy_eq_2} we obtain
\begin{align*}
b_n = \ell_{n,\alpha}
-\frac{2\ln(2\sqrt{\pi}|\alpha|)+\ln\ln n +\ln\Phi\left(\tau/\bar{\alpha}\right)-\tau^2/2}
{2\ell_{n,\alpha}}-\frac{\alpha\tau}{1 + \alpha^2} 
+ o(1/\ell_{n,\alpha})
 = \beta_n + o(\alpha_n).
\end{align*}
Finally, as in both cases $\alpha\geq 0$ and $\alpha<0$ we have
$a_n/\alpha_n\to 1$ and $(\alpha_n-\beta_n)/a_n\to 0$ as $n\to\infty$, then
by \citet[][Proposition 0.2]{resnick1987} we have $\Phi_{\alpha,\tau}(\alpha_n x+\beta_n)\to G_0(x)$
as $n\to\infty$.

%
%
%
%%%%%%%%%%%% END PROOF PROPOSITION 4 %%%%%%%%%%%%%%%

%%%%%%%%%%%% START PROOF THEOREM 1 %%%%%%%%%%%%%%%%
%
%
%
\subsection{Proof of Theorem~\ref{theo:RateAlphaBeta}}
\label{app:RateAlphaBeta}

Let $u_n = \alpha_n x + \beta_n$ and $v_n = \alpha_n x + \beta_n$ where $\alpha_n$ and 
$\beta_n$ are respectively defined as in Section~\ref{sec:AlphaBeta} when $\alpha \geq 0$ and 
$\alpha < 0$.
Let $\bar{\alpha}=(1+\alpha^2)^{1/2}$ for any $\alpha\in\real$.
When $\alpha \geq0$ it is easy to check that $u_n^{-2} \sim \ell_{n,0}^{-2}$ as $n\to\infty$, which implies that 
$O(u_n^{-2}) = O((\ln n)^{-1})$, and  
$$
u_n^{-1} = \ell_{n,0}^{-1} \left\{ 1 + O\left( \frac{\ln\ln n}{\ln n} \right) \right\}.
$$
Furthermore, by Proposition~\ref{prop:MillsIneq}\ref{en:first_cond_palpha_parg} we can write
$$
1- \Phi_{\alpha, \tau}(u_n) = u_n^{-1} \phi_{\alpha,\tau}(u_n) \{1 + O(u_n^{-2})\}.
$$
Defining $\vartheta_n = n\{1-\Phi(u_n; \alpha, \tau)\}$, we may then write
$$\vartheta_n = e^{-x} \left[ 1 - \frac{(\ln\ln n)^2}{16 \ln n}\{1+o(1)\}\right].$$
Setting $\vartheta = e^{-x}$ then gives
$$\vartheta - \vartheta_n = e^{-x} \frac{(\ln\ln n)^2}{16 \ln n}\{1+o(1)\}.$$
In the case when $\alpha <0$ it is also easy to check that 
$$
v_n^{-2} = \ell_{n,0}^{-2} \bar{\alpha}^2 
\left[ 1+ O\left\{ \left( \frac{\ln\ln n}{\ln n} \right)^2 \right\} \right]
$$
which implies that $O(v_n^{-2}) = O((\ln n)^{-1})$, and
$$
v_n^{-1} = \ell_{n,0}^{-1} \bar{\alpha} 
\left\{ 1 + O\left( \frac{\ln\ln n}{\ln n}\right) \right\}.
$$
Furthermore, by Proposition~\ref{prop:MillsIneq}\ref{en:second_cond_nalpha_narg} we have
$$
1- \Phi_{\alpha, \tau}(v_n) 
= -\phi(x) \phi(v_{\alpha ,\tau})
\left\{ \left( \bar{\alpha}v_n + \alpha \tau \right) v_{n, \alpha, \tau}  \Phi(\tau / \bar{\alpha}) \right\}^{-1}
\{1 + O(v_n^{-2})\},
$$
where $v_{n, \alpha, \tau} = \alpha v_n + \tau$.
Thus, when $\vartheta_n = n(1-\Phi(v_n; \alpha, \tau))$, using the additional approximation 
$v_n = \ell_{n,\alpha} + o(1)$, we can write
$$
\vartheta_n = e^{-x} \left[ 1 - \frac{(\ln\ln n)^2}{4 \ln n}\{1+o(1)\} \right],
$$
from which we obtain
$$
\vartheta - \vartheta_n = e^{-x} \frac{(\ln \ln n)^2}{4 \ln n}\{1+o(1)\}.
$$
Then apply \citet[Theorem~2.4.2]{leadbetter1983} to complete the proof of the first assertion 
of the theorem.
%
%
%
%%%%%%%%%%%% END PROOF THEOREM 1 %%%%%%%%%%%%%%%%%%

%%%%%%%%%%%% START PROOF THEOREM 2 %%%%%%%%%%%%%%%%
%
%
%

Focusing on the normalising constants $a_n$ and $b_n$ given in \eqref{eq:AnBn}, we require the following lemma to determine the speed of convergence to the Gumbel distribution.

%%% BEGIN SECOND LEMMA 

\begin{lem}
\label{lem:lim}
Let $h_{\alpha,\tau}(x;b_n) = n \log \Phi_{\alpha, \tau}(f(b_n) x + b_n) + e^{-x}$, where the 
normalising constant $b_n$ is given by \eqref{eq:AnBn} and $f$ is the auxiliary function 
defined in Proposition~\ref{prop:DomAtt}.
Then
$$
\lim_{n \rightarrow \infty} b_n^2 \left( b_n^2 h_{\alpha,\beta}(x; b_n) - \kappa(x)\right)
= \omega(x),
$$
where $\kappa(x)$ and $\omega(x)$ depend on the sign of the slant parameter $\alpha$ 
and are defined in the statement of Theorem~\ref{theo:RateAlphaBeta}.
\end{lem}
%

%%% BEGIN PROOF THEOREM 

The proof of Lemma \ref{lem:lim} is provided in the Supplementary Material.  Lemma~\ref{lem:lim} indicates that $h_{\alpha, \tau}(x, b_n ) \rightarrow 0$ as $n$ 
gets large and
$$
\Bigg|
\sum_{i=3}^\infty \frac{h^{i-3}_{\alpha, \tau}(x, b_n)}{i!}
\Bigg|
< \exp \left( h_{\alpha, \tau}(x, b_n) \right)
\rightarrow 1, \quad n\to\infty.
$$
Then, noting that
$\exp \left\{ h_{\alpha, \tau}(x, b_n) \right\} 
= \Phi^n_{\alpha, \tau}(a_n x + b_n) G_0(x)^{-1}$ 
 and applying Lemma~\ref{lem:lim} once more, we have 
\begin{align*}
&b_n^2 \left[ 
b_n^2 \left( \Phi^n_{\alpha, \tau}(a_n x + b_n) - G_0(x) \right)
- \kappa(x) G_0(x)
 \right]=
b_n^2 \left[ 
b_n^2 \left( \frac{\Phi^n_{\alpha, \tau}(a_n x + b_n)}{G_0(x)}-1 \right) - \kappa(x)
\right]
G_0(x) \\
&=
b_n^2 \left[ 
b_n^2 \left( \exp \left\{ h_{\alpha, \tau}(x, b_n) \right\} -1 \right) - \kappa(x)
\right]
G_0(x) \\
&=
b_n^2 \left[ 
b_n^2 \left( h_{\alpha, \tau}(x, b_n) 
+ \frac{h_{\alpha, \tau}(x, b_n)^2}{2} 
+ \sum_{i=3}^\infty \frac{h_{\alpha, \tau}(x, b_n)^i}{i!}
\right) - \kappa(x)
\right]
G_0(x) \\
&=
\left[ 
b_n^2 \left( b_n^2  h_{\alpha, \tau}(x, b_n) - \kappa(x) \right)
+ b_n^4 h_{\alpha, \tau}(x, b_n)^2
\left( 
\frac{1}{2} 
+ h(x, b_n; \alpha, \tau) 
\sum_{i=3}^\infty \frac{h_{\alpha, \tau}(x, b_n)^{i-3}}{i!}
\right)
\right]
G_0(x) \\
& \rightarrow 
\left[ 
\omega(x) + \frac{\kappa^2(x)}{2}
\right] G_0(x), \quad n \rightarrow \infty.
\end{align*}
%

%
%
%
%%%%%%%%%%%% END PROOF THEOREM 2 %%%%%%%%%%%%%%%%%%

%
\bibliographystyle{chicago}
\bibliography{biblio}

\begin{thebibliography}{}

\bibitem[\protect\citeauthoryear{Arellano-Valle and Genton}{Arellano-Valle and
  Genton}{2010}]{arellano2010}
Arellano-Valle, R.~B. and M.~G. Genton (2010).
\newblock Multivariate extended skew-{$t$} distributions and related families.
\newblock {\em Metron\/}~{\em 68\/}(3), 201--234.

\bibitem[\protect\citeauthoryear{Azzalini}{Azzalini}{1985}]{azzalini1985}
Azzalini, A. (1985).
\newblock A class of distributions which includes the normal ones.
\newblock {\em Scand. J. Statist.\/}~{\em 12\/}(2), 171--178.

\bibitem[\protect\citeauthoryear{Azzalini and Capitanio}{Azzalini and
  Capitanio}{2014}]{azzalini2014}
Azzalini, A. and A.~Capitanio (2014).
\newblock {\em The Skew-Normal and Related Families}.
\newblock Cambridge: University Press, Cambridge.

\bibitem[\protect\citeauthoryear{Beirlant, Goegebeur, Teugels, and
  Segers}{Beirlant et~al.}{2004}]{beirlant2004}
Beirlant, J., Y.~Goegebeur, J.~Teugels, and J.~Segers (2004).
\newblock {\em Statistics of Extremes: Theory and Applications}.
\newblock John Wiley \& Sons, Ltd., Chichester.

\bibitem[\protect\citeauthoryear{Leadbetter, Lindgren, and
  Rootz\'en}{Leadbetter et~al.}{1983}]{leadbetter1983}
Leadbetter, M.~R., G.~Lindgren, and H.~Rootz\'en (1983).
\newblock {\em Extremes and Related Properties of Random Sequences and
  Processes}.
\newblock Springer-Verlag, New York-Berlin.

\bibitem[\protect\citeauthoryear{Liao, Peng, Nadarajah, and Wang}{Liao
  et~al.}{2014}]{liao2014}
Liao, X., Z.~Peng, S.~Nadarajah, and X.~Wang (2014).
\newblock Rates of convergence of extremes from skew-normal samples.
\newblock {\em Statistics \& Probability Letters\/}~{\em 84}, 40 -- 47.

\bibitem[\protect\citeauthoryear{Mills}{Mills}{1926}]{mills1926}
Mills, J.~P. (1926).
\newblock Table of the ratio: Area to bounding ordinate, for any portion of
  normal curve.
\newblock {\em Biometrika\/}~{\em 18\/}(3/4), 395--400.

\bibitem[\protect\citeauthoryear{Resnick}{Resnick}{1987}]{resnick1987}
Resnick, S.~I. (1987).
\newblock {\em Extreme Values, Regular Variation, and Point Processes}.
\newblock Springer-Verlag.

\end{thebibliography}


\begin{thebibliography}{}

\bibitem[\protect\citeauthoryear{Canto~e Castro}{Canto~e
  Castro}{1987}]{castro1987}
Canto~e Castro, L. (1987).
\newblock Uniform rates of convergence in extreme-value theory---normal and
  gamma models.
\newblock {\em Ann. Sci. Univ. Clermont-Ferrand II Probab. Appl.\/}~(6),
  25--41.

\end{thebibliography}

\end{document}

\title{Supplementary material for `Extremal properties of the univariate extended skew-normal distribution'}

\author{B. Beranger\footnote{School of Mathematics and Statistics, University of New South Wales, Sydney, Australia.}\;\,\footnote{Communicating Author: {\tt B.Beranger@unsw.edu.au}},\; S. A. Padoan\footnote{Department of Decision Sciences, Bocconi University of Milan, Italy.
},\; Y. Xu$^*$\; and\; S. A. Sisson$^*$
}

\date{}

\maketitle
\begin{abstract}
This document contains the proof of Lemma 1 used in proof of Theorem 2.1 in the main paper.
\end{abstract}

Note that all references below of the form ($\cdot^*$) refer to equation ($\cdot$) in the main paper.
When we refer to a proposition or theorem we implicitly refer to a result in the paper unless otherwise specified. 
%
%
%
%%%%%%%%%%%% END PROOF PROPOSITION 4 %%%%%%%%%%%%%%%

%%%%%%%%%%%% START PROOF THEOREM 1 %%%%%%%%%%%%%%%%
%
%
%
\section{Technical details}
\label{app:details}

We first recall the statement.

\begin{lem}
\label{lem:lim}
Let $h_{\alpha,\tau}(x;b_n) = n \log \Phi_{\alpha, \tau}(f(b_n) x + b_n) + e^{-x}$, where the 
normalising constant $b_n$ is given by ($5^*$) and $f$ is the auxiliary function 
defined in Proposition~2.3.
Then,
$$
\lim_{n \rightarrow \infty} b_n^2 \left( b_n^2 h_{\alpha,\beta}(x; b_n) - \kappa(x)\right)
= \omega(x),
$$
where $\kappa(x)$ and $\omega(x)$ depend on the sign of the slant parameter $\alpha$ 
and are defined in Theorem~2.1.
\end{lem}
\begin{proof}
We first need the following auxiliary result.
\begin{lem}
\label{lem:approxTail}
For large $x$, the tail probability associated with the univariate extended skew-Normal distribution
$ESN(\alpha, \tau)$ is given, as a function of the sign of the slant parameter $\alpha$, by
\begin{inparaenum} 
\item For $\alpha\geq 0$, 
$$
1-\Phi_{\alpha, \tau}(x) = (2\pi e)^{-1/2} \left( 1 - x^{-2} + 3 x^{-4} + O(x^{-6})\right) 
\frac{\Phi( x_{\alpha, \tau} )}{\Phi(\sqrt{\tau_\alpha \tau / \alpha})}
\exp \left( - \int_1^x  \frac{g(t)}{f(t)} \der t \right).
$$
\item For $\alpha < 0$
\begin{align*}
1 - \Phi_{\alpha, \tau}(x) &= 
\frac{-\exp \left\{ - \frac{1 + (\alpha + \tau)^2}{2}\right\}}
{2\pi \Phi(\sqrt{\tau_\alpha \tau / \alpha}) (\alpha(\alpha +\tau)+1)(\alpha + \tau) } 
\times\exp \left\{ -\int_1^x 
\frac{g(t)}{f(t)} \der t
\right\} \\
& \quad \times \Bigg\{
\left( 1-\frac{1}{x^2} + \frac{3}{x^4}\right)
\left( 1 - \frac{\alpha^2 \tau_\alpha}{x} - \frac{\alpha \tau \tau_\alpha}{x^2} \right)
- \left( 1-\frac{1}{x^2} - \frac{3}{x_{\alpha, \tau}^2}\right) 
 \left( \frac{\bar{\alpha}^2}{x_{\alpha, \tau}^2} + \frac{\alpha \tau}{x x_{\alpha, \tau}^2}\right)\\
& \qquad\quad 
+ \frac{\alpha x_{\alpha, \tau} }{\bar{\alpha}^2}
\Bigg[
\frac{\alpha \tau x + 2}{x^3} 
- \frac{\alpha \tau x + 4}{x^5} 
- \frac{\alpha \tau x + 2}{x^3 \left( x + \tau_\alpha \right)^2 \bar{\alpha}^2} 
- \frac{2\alpha \tau x + 6}{x^4 \left( x+ \tau_\alpha \right) \bar{\alpha}^2} +\frac{3\alpha \tau}{x^6}\\
& \left. \left. \qquad\quad
- \frac{\tau_\alpha^2 (\alpha^2 -1)}
{x^4 \left( x + \tau_\alpha \right)^2  \tau \alpha}
+ \frac{4 \tau_\alpha }
{x^5 \left( x + \tau_\alpha \right)} 
+ \frac{11 \tau_\alpha^2}
{x^3 \left( x + \tau_\alpha \right)^3 \alpha \tau}
+ \frac{6 \tau_\alpha^2}
{x^2 \left( x + \tau_\alpha \right)^4 \alpha \tau}
\right] 
+ O\left( \frac{1}{x^6} \right)
\right\},
\end{align*}
\end{inparaenum}
where $\tau_\alpha = \tau \alpha / \bar{\alpha}^2, \bar{\alpha} + \alpha \tau >0, \alpha + \tau <0 $  and $f(t)$ and $g(t)$ are defined in Proposition~2.3.
\end{lem}
%
%%% END SECOND LEMMA
%
%%%%%%%%%%%%%%%%%%%%%%%%%%%%%%%%%%%%%%%%%%%%%
%
% Proof of Lemma 2
%
%%%%%%%%%%%%%%%%%%%%%%%%%%%%%%%%%%%%%%%%%%%%%
%
\begin{proof}

%%% BEGIN FIRST LEMMA 

The tail probability can be written as
\begin{align*}
1-\Phi_{\alpha,\tau}(x) = 
\int_x^\infty \frac{1}{t} 
\left( 
t \phi_{\alpha, \tau}(t) 
+ \frac{ \tau_\alpha (\alpha \tau t + 1)}{ \tau t^2}
\frac{\phi(t)\phi( t_{\alpha,\tau})}{\Phi(\sqrt{\tau_\alpha \tau / \alpha})} 
\right) \der t 
 - \int_x^\infty 
\frac{\tau_\alpha (\alpha\tau t + 1)}{ \tau t^3}
\frac{\phi(t)\phi(t_{\alpha,\tau})}{\Phi(\sqrt{\tau_\alpha \tau / \alpha})}
\der t,
\end{align*}
which from the change of variables
$$
u = - t^{-1},
\quad \der v = - t \phi_{\alpha, \tau}(t)
- \frac{\tau_\alpha (\alpha \tau t + 1)}{\tau t^2}
\frac{\phi(t)\phi( t_{\alpha, \tau})}{\Phi(\sqrt{\tau_\alpha \tau / \alpha})}, 
\quad \der u = t^{-2},
\quad v 
= \phi_{\alpha, \tau}(t) + \frac{\alpha}{(1+\alpha^2)t}
\frac{\phi(t)\phi(t_{\alpha, \tau})}{\Phi(\sqrt{\tau_\alpha \tau / \alpha})},
$$
yields
\begin{align*}
1-\Phi_{\alpha,\tau}(x) =& 
\frac{1}{x} \left( 
\phi_{\alpha, \tau}(x) + \frac{\tau_\alpha}{\tau x}
\frac{\phi(x)\phi( t_{\alpha, \tau})}{\Phi(\sqrt{\tau_\alpha \tau / \alpha})}
\right)
- \int_x^\infty \frac{1}{t^2} \phi_{\alpha, \tau}(t) \der t 
- \int_x^\infty \frac{ \tau_\alpha (\alpha\tau t + 2)}{ \tau t^3 }
\frac{\phi(t)\phi(t_{\alpha, \tau})}{\Phi(\sqrt{\tau_\alpha \tau / \alpha})} \der t.
\end{align*}
Applying a similar change of variable to the remaining two integrals and iterating these steps multiple times gives
\begin{align}
\label{eq:TailExpansion}
1-\Phi_{\alpha, \tau}(x) &=
\frac{\phi_{\alpha, \tau}(x)}{x} 
\Bigg\{
1-\frac{1}{x^2}+\frac{3}{x^4}-\frac{15}{x^6} 
+ \frac{\phi( x_{\alpha, \tau})}{\Phi( x_{\alpha, \tau})}
\Bigg[ 
\frac{\tau_\alpha}{x \tau}
- \frac{\tau_\alpha}{x^3 \tau} 
- \frac{\tau_\alpha^2 (\alpha\tau x + 2)}{x^2 ( x + \tau_\alpha) \alpha \tau^2}
+ \frac{3 \tau_\alpha}{x^5 \tau}+ \frac{\tau_\alpha^2 (\alpha\tau x + 4)}
{x^4 ( x + \tau_\alpha )\alpha \tau^2 }
\nonumber\\
& \qquad\qquad\qquad
+ \frac{\tau_\alpha^3 (\alpha\tau x + 2)}
{x^2 ( x + \tau_\alpha )^3 \alpha^2 \tau^3 }
+ \frac{\tau_\alpha^3 (2\alpha\tau x + 6)}
{x^3 ( x + \tau_\alpha )^2 \alpha^2 \tau^3 } 
- \frac{15 \tau_\alpha}{x^7 \tau}
- \frac{\tau_\alpha^2 (3\alpha\tau x + 18)}
{x^6 ( x + \tau_\alpha ) \alpha \tau^2 }
- \frac{\tau_\alpha^3 (\alpha\tau x + 4)}
{x^4 ( x + \tau_\alpha )^3 \alpha^2 \tau^3 }
\nonumber\\
& \qquad\qquad\qquad 
- \frac{ \tau_\alpha^3 (4\alpha\tau x + 20)}
{x^5 ( x + \tau_\alpha )^2 \alpha^2 \tau^3 }
+ \frac{\tau_\alpha^4 }
{x^2 ( x + \tau_\alpha )^4 \alpha^2 \tau^3 } 
+ \frac{2 \tau_\alpha^4 }
{x^3 ( x + \tau_\alpha)^3 \alpha^2 \tau^3 }
- \frac{\tau_\alpha^4 (\alpha\tau x + 2) ( 6x + 3 \tau_\alpha)}
{x^3 ( x + \tau_\alpha)^5 \alpha^3 \tau^4 } 
\nonumber\\
& \qquad\qquad\qquad
- \frac{ \tau_\alpha^4 (2\alpha\tau x + 6) ( 6x + 4 \tau_\alpha)}
{x^4 ( x + \tau_\alpha )^4 \alpha^3 \tau^4 }
\Bigg]
\Bigg\} 
+ R,
\end{align}
where
\begin{align*}
R &= 
\int_x^\infty \frac{105}{t^8} \phi_{\alpha,\tau}(t) \der t
+ \int_x^\infty \frac{15 \tau_\alpha (\alpha \tau t + 8)}{t^9 \tau}
\frac{\phi(t)\phi( t_{\alpha,\tau})}{\Phi(\sqrt{\tau_\alpha \tau / \alpha})} \der t \\
&\quad + \int_x^\infty \frac{\tau_\alpha^2}{\alpha \tau^2} 
\left[ 
\frac{3\alpha \tau}{t^7 ( t + \tau_\alpha )} 
+ \frac{(3\alpha \tau t + 18)\left( 8t + 7\tau_\alpha \right)}
{t^8 \left( t + \tau_\alpha \right)^2}
\right]
\frac{\phi(t)\phi( t_{\alpha,\tau})}{\Phi(\sqrt{\tau_\alpha \tau / \alpha})} \der t \\
&\quad - \int_x^\infty \frac{\tau_\alpha^3}{\alpha^2 \tau^3} 
\left[ 
 \frac{\alpha \tau}{t^5 \left( t + \tau_\alpha \right)^3}
+ \frac{4\alpha \tau}{t^6 \left( t + \tau_\alpha \right)^2}
- \frac{(\alpha \tau t + 4) \left( 8t + 5\tau_\alpha \right) }
{t^6 \left( t + \tau_\alpha \right)^4} 
- \frac{(4 \alpha \tau t + 20) \left( 8t + 6 \tau_\alpha \right) }
{t^7 \left( t + \tau_\alpha \right)^3}
\right]
\frac{\phi(t)\phi( t_{\alpha,\tau})}{\Phi(\sqrt{\tau_\alpha \tau / \alpha})} \der t \\
&\quad - \int_x^\infty \frac{\tau_\alpha^4}{\alpha^3 \tau^4}
\left[
\frac{\alpha \tau \left( 19t + 6 \tau_\alpha \right) }
{t^4 \left( t + \tau_\alpha \right)^5}
+ \frac{2\alpha \tau \left( 19t + \tau_\alpha \right) +36}
{t^4 \left( t + \tau_\alpha \right)^5} 
-\frac{(\alpha\tau t + 2) \left( 6t + 3 \tau_\alpha \right) 
\left( 9t + 4 \tau_\alpha \right)}
{t^5 \left( t +\tau_\alpha \right)^6} 
\right. \\
& \qquad\qquad\qquad \quad \left.
-\frac{(2\alpha\tau t + 6) \left( 6t + 4 \tau_\alpha \right) 
\left( 9t + 5 \tau_\alpha \right)}
{t^6 \left( t + \tau_\alpha \right)^5}
\right]
\frac{\phi(t)\phi( t_{\alpha,\tau})}{\Phi(\sqrt{\tau_\alpha \tau / \alpha})} \der t.
\end{align*}
Through similar calculations it can be shown that
\begin{align*}
\int_x^\infty \frac{1}{t^8} \phi(t;\alpha,\tau) \der t
< \frac{1}{x^9}\phi(x;\alpha,\tau) 
+ \int_x^\infty \frac{\alpha}{t^9} \frac{\phi(t)\phi(\alpha t + \tau)}
{\Phi(\tau/\sqrt{1+\alpha^2})} \der t,
\end{align*}
and then it is trivial to prove that each term of $R$ converges to $0$ with rate $o(x^{-8})$ or 
faster.\\
When $\alpha\geq0$, then $x^5 \phi(\alpha x + \tau) \rightarrow 0$ and 
$\Phi(\alpha x + \tau) \rightarrow 1$, as $n \rightarrow \infty$
and thus \eqref{eq:TailExpansion} becomes
 \begin{align*}
 1-\Phi_{\alpha, \tau}(x) 
&= \frac{1}{\sqrt{2\pi e}} \left\{ 1-\frac{1}{x^2}+\frac{3}{x^4} 
+ O\left(\frac{1}{x^6}\right) \right\}
\frac{\Phi( x_{ \alpha, \tau})}{\Phi( \sqrt{\tau_\alpha \tau / \alpha} )}
\exp \left( - \int_x^\infty \frac{g(t)}{f(t)} \der t \right).
\end{align*}
When $\alpha <0$, then 
using the results in \citet{castro1987} we have that
\begin{align*}
\Phi(x_{\alpha, \tau})\left\{ 1-\frac{1}{x^2}+\frac{3}{x^4} + O\left(\frac{1}{x^6}\right)\right\}
&= \frac{-\phi(x_{\alpha, \tau})}{x_{\alpha, \tau}}
\left(
1 - \frac{1}{x^2} - \frac{1}{x_{\alpha,\tau}^2} + \frac{3}{x^4} 
+ \frac{1}{x^2 x_{\alpha, \tau}}
+ \frac{3}{x_{\alpha, \tau}^4} + O\left( \frac{1}{x^6} \right)
\right),
\end{align*}
then we have
\begin{align*}
1 - \Phi_{\alpha, \tau}(x) &= 
\frac{-\phi(x)\phi(x_{\alpha, \tau}) f(x)}
{ x_{\alpha, \tau} 
\Phi(\sqrt{\tau_\alpha \tau / \alpha})} \\
& \times \Bigg\{
\left( 1-\frac{1}{x^2} + \frac{3}{x^4}\right)
\left( 1 - \frac{\alpha^2 \tau_\alpha}{x} - \frac{\alpha \tau \tau_\alpha}{x^2}\right)
- \left( 1-\frac{1}{x^2} - \frac{3}{ x_{\alpha, \tau}^2}\right)
\left( \frac{\alpha^2 +1}
{x_{\alpha, \tau}^2} + \frac{\alpha \tau}{x x_{\alpha, \tau}^2}\right)
 \\
& \qquad
+ \frac{\tau_\alpha x_{\alpha, \tau}}{\tau}
\Bigg[
\frac{\alpha \tau x + 2}{x^3}  
- \frac{\alpha \tau x + 4}{x^5}
- \frac{\tau_\alpha x + 2}{x^3 \left( x + \tau_\alpha \right)^2 }
- \frac{2 \tau_\alpha x + 6}{x^4 \left( x+ \tau_\alpha \right) }
+\frac{3\alpha \tau}{x^6}
- \frac{\tau_\alpha^2 (\alpha^2 -1)}
{x^4 \left( x + \tau_\alpha \right)^2 \alpha \tau}
\\
& \qquad\qquad\qquad\,
+ \frac{4 \tau_\alpha}
{x^5 \left( x + \tau_\alpha \right) } 
\left. \left.
+ \frac{11 \tau_\alpha}
{x^3 \left( x + \tau_\alpha \right)^3 \alpha \tau}
+ \frac{6 \tau_\alpha}
{x^2 \left( x + \tau_\alpha \right)^4 \alpha \tau}
\right] 
+ O\left( \frac{1}{x^6} \right)
\right\}.
\end{align*}
Finally the proof is completed by taking the following result from 
Section~A3, when $\alpha < 0$:
\begin{align*}
\frac{-\phi(x) \phi( x_{\alpha, \tau}) f(x)}
{ x_{\alpha, \tau} \Phi(\sqrt{\tau_\alpha \tau / \alpha})}
& = \frac{-e^{- \frac{1 + (\alpha + \tau)^2}{2}}}
{2\pi \Phi(\sqrt{\tau_\alpha \tau / \alpha}) \{\alpha(\alpha +\tau)+1\}(\alpha + \tau) }
\exp \left( -\int_1^x 
\frac{g(t)}{f(t)} \der t
\right).
\end{align*}
\end{proof}
%
%%% END SECOND LEMMA

When $\alpha \geq0$,
then by Proposition~2.3, the auxiliary function is 
$f(t)= 1/t$.
Then by equation  (5*) we have $1-\Phi_{\alpha, \tau}(b_n) = n^{-1}$
so that $b_n \rightarrow \infty$ as $n \rightarrow \infty$.
Using Proposition~2.2 and the approximation of 
$\phi_{\alpha, \tau}(b_n^{-1} x + b_n)$ it can be shown that
$$
\begin{array}{ccc}
\lim_{n \rightarrow \infty} 
b_n \frac{1-\Phi_{\alpha, \tau}(x_{b_n})}{\phi_{\alpha, \tau}(b_n)}
= \exp(-x)
& \quad \textrm{ and } 
&
\quad
\lim_{n \rightarrow \infty} 
\frac{1-\Phi_{\alpha, \tau} (x_{b_n})}{b_n^{-4}}
= 0,
\end{array}
$$
where $x_{b_n} = b_n^{-1} x + b_n$.
Now, define
$$
A_{\alpha, \tau}(b_n) = 
\frac{\Phi(b_{n,\alpha, \tau}) \left( 1 - b_n^{-2} + 3 b_n^{-4} + O(b_n^{-6})\right)}
{\Phi( x_{b_n, \alpha, \tau} ) \left( 1 - x_{b_n}^{-2} + 3 x_{b_n}^{-4} + O(x_{b_n}^{-6})\right)},
$$
with $b_{n,\alpha, \tau} = \alpha b_n + \tau$ and $x_{b_n, \alpha, \tau} = \alpha x_{b_n} + \tau$,
which has the  properties
$$
\begin{array}{ccc}
\lim_{n \rightarrow \infty} A_{\alpha, \tau}(b_n) = 1,
&
\lim_{n \rightarrow \infty} \frac{A_{\alpha, \tau}(b_n) - 1}{b_n^{-2}} = 0,
&
\lim_{n \rightarrow \infty} \frac{A_{\alpha, \tau}(b_n) - 1}{b_n^{-4}} = -2x.
\end{array}
$$
Applying Lemma~\ref{lem:approxTail}(i) gives
\begin{align*}
\frac{1-\Phi_{\alpha, \tau}(b_n)}{1-\Phi_{\alpha, \tau}(x_{b_n})} e^{-x}
&= A_{\alpha, \tau}(b_n)
\frac{\exp\left( - \int_1^{b_n} t \left( 1 + \frac{1}{t^2} \right) \der t \right)}
{\exp\left( - \int_1^{x_{b_n} } t \left( 1 + \frac{1}{t^2} \right) \der t \right)}
e^{-x} \\
&= A_{\alpha, \tau}(b_n)
\left\{
1 + \int_0^x  \frac{t}{b_n^2} + \frac{1}{b_n^2 + t } \der t
+ \frac{1}{2} \left( \int_0^x  \frac{t}{b_n^2} + \frac{1}{b_n^2 + t } \der t \right)^{1/2}
(1 + o(1))
\right\}.
\end{align*}
Using the above results leads to 
\begin{align*}
\lim_{n \rightarrow \infty}  \frac{h_{\alpha, \tau}(x; b_n)}{b_n^{-2}}
&= \lim_{n \rightarrow \infty} 
\frac{\log \Phi_{\alpha, \tau}( x_{b_n} ) + (1-\Phi_{\alpha, \tau}(b_n)) e^{-x}}
{b_n n^{-1} \times b_n^{-3}} \\
&= \lim_{n \rightarrow \infty} 
\frac{- \left( 1 - \Phi_{\alpha, \tau}( x_{b_n} )\right) 
- \frac{1}{2} \left( 1 - \Phi_{\alpha, \tau}( x_{b_n} )\right)^2 (1+o(1))} 
{\phi_{\alpha, \tau} (b_n) b_n^{-3}} 
 + \lim_{n \rightarrow \infty} 
\frac{\left( 1 - \Phi_{\alpha, \tau}(b_n)\right)} {\phi_{ \alpha, \tau}(b_n) b_n^{-3}} e^{-x} \\
& = \lim_{n \rightarrow \infty} 
\frac{b_n \left( 1 - \Phi_{\alpha, \tau}(x_{ b_n}) \right)}{\phi_{\alpha, \tau}(b_n)}
\frac{\frac{1 - \Phi_{\alpha, \tau}(b_n)}{1 - \Phi_{\alpha, \tau}(x_{b_n} )} e^{-x} -1}
{b_n^{-2}} \\
& \approx \lim_{n \rightarrow \infty} \frac{e^{-x}}{b_n^{-2}} 
\left\{ 
A_{\alpha, \tau}(b_n) - 1 + A_{\alpha, \tau}(b_n) 
\int_0^x \frac{t}{b_n^2} + \frac{1}{b_n^2 + t} \der t (1+o(1))
\right\}\\
&= e^{-x} \lim_{n \rightarrow \infty} \int_0^x t + \frac{1}{1+b_n^{-2}t} \der t \\
&= e^{-x} \left( \frac{x^2}{2} + x \right) = \kappa(x).
\end{align*}
In a similar way, we have
\begin{align*}
\lim_{n \rightarrow \infty} & b_n^2\left( b_n^2 h_{\alpha, \tau}(x; b_n) - \kappa(x) \right) \\
&= \lim_{n \rightarrow \infty} 
\frac{\log \Phi_{\alpha, \tau} (x_{b_n} ) + (1-\Phi_{\alpha, \tau}(b_n)) e^{-x} 
\left( 1 - b_n^{-2} \left( \frac{x^2}{2} + x \right) \right)}
{\phi_{\alpha, \tau}(b_n) b_n^{-5}} \\
&= \lim_{n \rightarrow \infty}
\frac{b_n \left( 1 - \Phi_{\alpha, \tau}( x_{b_n}) \right)}{\phi_{\alpha, \tau}(b_n)}
\times
\frac{\frac{1 - \Phi_{\alpha, \tau}(b_n)}{1 - \Phi_{\alpha, \tau} ( x_{b_n} )} e^{-x} 
\left( 1 - b_n^{-2} \left( \frac{x^2}{2} + x \right) \right) -1}
{b_n^{-4}} \\
& \approx e^{-x}  \lim_{n \rightarrow \infty} \left\{
A_{\alpha, \tau}(b_n) b_n^2 
\left( \int_0^x \frac{t}{b_n^2} + \frac{1}{b_n^2 + t } \der t - \frac{x^2 + 2x}{2}  \right)
\right. \\
& \qquad\qquad\qquad  - A_{\alpha, \tau}(b_n)
\left( \int_0^x \frac{t}{b_n^2} + \frac{1}{b_n^2 + t } \der t \right)^2
\left( \frac{1}{2} - \frac{x^2 + 2x}{4 b_n^2}\right) 
+ \frac{A_{\alpha, \tau}(b_n)-1}{b_n^{-4}} \bigg\} \\
& = e^{-x} \left( - \frac{x^2}{2} - \frac{1}{2}\left( \frac{x}{2} + x\right)^2 -2x \right) \\
& = -\frac{e^{-x}}{8} \left( x^4 + 4 x^3 + 8 x^2 + 16 x \right) = \omega(x).
\end{align*}
Now, consider $\alpha <0$.
By Proposition~2.3 the auxiliary function is 
$f(t) = 1/(\bar{\alpha}^2t + \alpha \tau)$. Using Proposition~2.2 and after some simple manipulations we obtain
$$
\begin{array}{ccc}
\lim_{n \rightarrow \infty} 
b_n \frac{1 - \Phi_{\alpha, \tau}(f(b_n)x + b_n)}{\phi_{\alpha, \tau}(b_n) (\alpha^2 +1)^{-1}}
= e^{-x}
& \quad \textrm{ and } 
& \quad \lim_{n\rightarrow \infty} \frac{1-\Phi_{\alpha, \tau}(f(b_n)x + b_n)}{b_n^{-4}}=0.
\end{array}
$$
Define
$$
B_{\alpha, \tau}(b_n) = 
\frac{C_{ \alpha, \tau}(b_n)}{C_{\alpha, \tau} \left( f(b_n) x + b_n\right)},
$$
where 
\begin{align*}
C_{\alpha, \tau}(x) &=
\left( 1-\frac{1}{x^2} + \frac{3}{x^4}\right)
\left( 1 - \frac{\alpha^2 \tau_\alpha}{x} - \frac{\alpha \tau \tau_\alpha}{x^2} \right)
- \left( 1-\frac{1}{x^2} - \frac{3}{x_{\alpha, \tau}^2}\right) 
\left( \frac{\alpha^2 +1}{x_{\alpha, \tau}^2} + \frac{\alpha \tau}{x x_{\alpha, \tau}^2}\right)\\
& \quad
+ \frac{x_{\alpha, \tau} \tau_\alpha}{\tau}
\Bigg[
\frac{\alpha \tau x + 2}{x^3}  
- \frac{\alpha \tau x + 4}{x^5}
- \frac{\alpha \tau x + 2}{x^3 \tau_\alpha^2 (1+\alpha^2)}
- \frac{2\alpha \tau x + 6}{x^4 \left( x+ \tau_\alpha \right) (1+\alpha^2)}
+\frac{3\alpha \tau}{x^6}
\\
& \qquad\qquad\qquad
- \frac{ \tau_\alpha^2 (\alpha^2 -1)}
{x^4 \left( x + \tau_\alpha \right)^2 \alpha \tau}
+ \frac{4 \tau_\alpha}
{x^5 \left( x + \tau_\alpha \right) } 
+ \frac{11 \tau_\alpha^2}
{x^3 \left( x + \tau_\alpha\right)^3 \alpha \tau}
+ \frac{6 \tau_\alpha^2}
{x^2 \left( x + \tau_\alpha\right)^4 \alpha \tau}
\Bigg]
+ O\left( \frac{1}{x^6} \right). 
\end{align*}
Clearly $\lim_{x \rightarrow \infty} C_{\alpha, \tau}(x)=1$ and thus 
$\lim_{n \rightarrow \infty} B_{\alpha, \tau}(b_n)=1$.
After some calculations it can then be shown that
\begin{align*}
B_{\alpha, \tau}(b_n) -1 
= f(b_n) \left\{ 
-2x b_n^{-3}
\left( \frac{3\alpha^2 + 1}{\alpha^2(\alpha^2 +1)} \right)
+ 3x \tau b_n^{-4}
\left( \frac{6 \alpha^4 + 5 \alpha^5 + 2}{\alpha^3 (\alpha^2 +1)^2} \right)
+ O\left(\frac{1}{b_n^6}\right)
\right\}
\end{align*}
and consequently
$$
\begin{array}{ccc}
\lim_{n \rightarrow \infty} \frac{B_{\alpha, \tau}(b_n) -1 }{b_n^{-2}} =0
& \quad \textrm{ and }
& \quad \lim_{n \rightarrow \infty} \frac{B_{\alpha, \tau}(b_n) -1 }{b_n^{-4}}
= \frac{-2x (1+ 3\alpha^2)}{\alpha^2 (\alpha^2 +1)^2}.
\end{array}
$$
Thus, using Lemma~\ref{lem:approxTail}(ii) and similar calculations as in Section A3,
we can now write
\begin{align*}
\frac{1-\Phi_{\alpha, \tau}(b_n)}{1-\Phi_{\alpha, \tau}(f(b_n) x + b_n)} e^{-x}
&= B_{\alpha, \tau}(b_n)
\frac{\exp \left\{ - \int_1^{b_n} \frac{g(t)}{f(t)} \der t \right\}}
{\exp \left\{ - \int_1^{f(b_n) x + b_n} \frac{g(t)}{f(t)} \der t \right\}}
e^{-x}\\
&= B_{\alpha, \tau}(b_n)
\exp \left\{ - \int_0^x \frac{g(f(b_n)t + b_n)}{f(f(b_n)t + b_n)} f(b_n) -1 \der t \right\},
\end{align*}
where
\begin{align*}
\frac{g(f(b_n)t + b_n)}{f(f(b_n)t + b_n)} f(b_n) -1
&=  t (\alpha^2 +1) f(b_n)^2
+ \frac{\alpha^2 + 1}{f(b_n)^2 + (\alpha^2 + 1)t}
+ \frac{\alpha}{ b_{n, \alpha, \tau} / f(b_n) + \alpha \tau}
\equiv D_{\alpha, \tau}(t, b_n).
\end{align*}
Consequently we have
\begin{align*}
\frac{1-\Phi_{\alpha, \tau}(b_n)}{1-\Phi_{\alpha, \tau}(f(b_n) x + b_n)} e^{-x}
= B_{\alpha, \tau}(b_n) \left\{ 1 + \int_0^x D_{\alpha, \tau}(t, b_n) \der t 
+ \frac{1}{2} \left( \int_0^x D_{\alpha, \tau}(t, b_n) \der t \right)^2 \right\}.
\end{align*}
Then, using the above results we have
\begin{align*}
\lim_{n \rightarrow \infty} b_n^2 h_{\alpha, \tau}(x; b_n)
&= \lim_{n \rightarrow \infty} 
\frac{\log \Phi_{\alpha, \tau} (f(b_n) x + b_n) + (1-\Phi_{\alpha, \tau}(b_n)) e^{-x}}
{\phi_{\alpha, \tau}(b_n) b_n^{-2} f(b_n)} \\
&= \lim_{n \rightarrow \infty} b_n 
\frac{(1-\Phi_{\alpha, \tau}(b_n))}{\phi_{\alpha, \tau}(b_n) (\alpha^2 + 1)^{-1}}
\times \frac{\frac{1-\Phi_{\alpha, \tau}(b_n)}{1-\Phi_{\alpha, \tau}(f(b_n) x + b_n)} e^{-x}-1}
{b_n^{-1}}
\frac{ (\alpha^2 + 1)^{-1} }{ f(b_n) }\\
&= e^{-x} \lim_{n \rightarrow \infty} 
\frac{ B_{ \alpha, \tau}(b_n) \left\{ 1 + \int_0^x D_{\alpha, \tau}(t, b_n) \der t 
\right\} -1}
{b_n} 
\frac{ (\alpha^2 + 1)^{-1} }{ f(b_n) } \\
&= e^{-x} \lim_{n \rightarrow \infty} 
\frac{B_{\alpha, \tau}(b_n) - 1 + B_{\alpha, \tau}(b_n) \int_0^x D_{\alpha, \tau}(t, b_n) \der t }
{b_n^{-2}} \\
& = e^{-x} \int_0^x \frac{t + 2}{\alpha^2 + 1} \der t \\
&= e^{-x} \frac{x^2 + 4x}{2(\alpha^2 + 1)}
= \kappa(x).
\end{align*}
Through similar calculations we obtain
\begin{align*}
\lim_{n \rightarrow \infty} & b_n^2\left( b_n^2 h_{\alpha, \tau}(x; b_n) - \kappa(x) \right) \\
&= \lim_{n \rightarrow \infty} 
\frac{\log \Phi_{\alpha, \tau} (x_{b_n} ) + (1-\Phi_{\alpha, \tau}(b_n)) e^{-x} 
\left( 1 - b_n^{-2} \left( \frac{x^2 + 4x}{2(\alpha^2 + 1 )} \right) \right)}
{\phi_{\alpha, \tau}(b_n) b_n^{-4} ((\alpha^2 +1) b_n + \alpha \tau)^{-1}} \\
&= \lim_{n \rightarrow \infty}
\frac{b_n \left( 1 - \Phi_{\alpha, \tau}( x_{b_n}) \right)}
{\phi_{\alpha, \tau}(b_n) (1 + \alpha^2)^{-1}}
\times
\frac{\frac{1 - \Phi_{\alpha, \tau}(b_n)}{1 - \Phi_{\alpha, \tau} ( x_{b_n} )} e^{-x} 
\left( 1 - b_n^{-2} \left( \frac{x^2 + 4x}{2(\alpha^2 + 1 )} \right) \right) -1}
{b_n^{-3}} 
\times \frac{(1 + \alpha^2)^{-1}}{((1 + \alpha^2) b_n + \alpha \tau)^{-1}}
\\
& \approx e^{-x}  \lim_{n \rightarrow \infty} \left\{
B_{\alpha, \tau}(b_n) b_n^2 
\left( \int_0^x D_{\alpha, \tau}(t, b_n) \der t - \frac{x^2 + 4x}{2(\alpha^2 +1)}  \right)
\right. \\
& \qquad\qquad\qquad  - A_{\alpha, \tau}(b_n)
\left( \int_0^x D_{\alpha, \tau}(t, b_n) \der t \right)^2
\left( \frac{1}{2} - \frac{x^2 +4x}{4 (\alpha^2 + 1) b_n^2}\right) 
+ \frac{B_{\alpha, \tau}(b_n)-1}{b_n^{-4}} \bigg\} \\
& = - e^{-x} \left( \frac{x^2}{(1 + \alpha^2)^2} 
+ \frac{1}{2} \left( \frac{x^2 + 4 x}{2 (1 + \alpha^2)} \right)^2 
+ \frac{2x (1 + 3 \alpha^2)}{\alpha^2 (\alpha^2 + 1)^2}
\right) \\
& = - e^{-x} \frac{\alpha^2 (1 + \alpha^2)^2}{8} 
\left\{ (1 + 3 \alpha^2) 16 x + \alpha^2 (x^4 + 8x^3 + 24 x^2) \right\}
= \omega(x)
\end{align*}
and the proof is complete.
%%% END FIRST LEMMA

\end{proof}

\bibliographystyle{chicago}
\bibliography{biblio}